\documentclass[bibnotes, amsmath,amssymb,
aps,
prx,
superscriptaddress,
twocolumn
]{revtex4-2}

\usepackage{graphicx}
\usepackage{dcolumn}
\usepackage{bm}

\usepackage{tikz}
\usetikzlibrary{arrows, chains, matrix, patterns, mindmap, calc, shapes.geometric}
\usepackage{subcaption}
\usepackage{enumitem}

\usepackage{listings}

\usepackage[subpreambles=true]{standalone}
\usepackage{import}
\usepackage[toc,page]{appendix}

\usepackage{physics}
\usepackage{mathtools}
\usepackage{bbm}
\usepackage{hyperref}

\usepackage{dsfont}

\newcommand{\methodfull}[0]{Non-Markovian Ensemble Propagation}
\newcommand{\method}[0]{NMEP}

\begin{document}

\preprint{APS/123-QED}

\title{\methodfull{}}

\author{Miralem Sinanović}
\affiliation{Theoretical Physics, Universität des Saarlandes, 66123 Saarbrücken, Germany}
\affiliation{Institute for Quantum Computing Analytics (PGI-12), Forschungszentrum Jülich, 52425 Jülich, Germany}
\author{Alessandro Ciani}
\affiliation{Institute for Quantum Computing Analytics (PGI-12), Forschungszentrum Jülich, 52425 Jülich, Germany}
\author{Shai Machnes}
\affiliation{Qruise GmbH, 66113 Saarbrücken, Germany}
\author{Frank K. Wilhelm}
\affiliation{Theoretical Physics, Universität des Saarlandes, 66123 Saarbrücken, Germany}
\affiliation{Institute for Quantum Computing Analytics (PGI-12), Forschungszentrum Jülich, 52425 Jülich, Germany}
\thanks{Qruise GmbH, 66113 Saarbrücken, Germany}

\date{\today}

\begin{abstract}

Open quantum systems are ubiquitous in nature and central to quantum technologies. A common description of their dynamics is given by the celebrated Lindblad master equation, which can be generalized to the non-Markovian scenario. 
In this work, we introduce the \methodfull{} (\method{}) method, which extends the Monte Carlo Wave-Function (MCWF) method to the non-Markovian case in a simple and general manner. 
We demonstrate its accuracy and effectiveness in a selection of examples, and compare the results with either analytic expressions or direct numerical integration of the master equation.
\end{abstract}

\maketitle

\section{Introduction}
The foundational dynamical equation of quantum physics is the well-known Schrödinger equation. It describes the evolution of state vectors, which, according to the axioms of quantum mechanics, give a complete description of the state of a closed quantum system. However, realistic systems interact with the environment and are accessible at least to the experimenter or observer. Thus, it is necessary to regard them as open quantum systems. The general idea of the theory of open quantum systems is to determine the dynamics of the system of interest by describing the environment, which we also call reservoir or bath depending on the context, using its statistical properties. There is a wealth on methods for implementing this concept, of which master equations, i.e., equations of motion for the density matrix $\rho$,  are arguably the most popular. 

The master equations, that are usually derived starting from a closed system description of the system of interest and the environment, differ in mathematical structure and complexity, and a concrete choice depends on the system and application at hand, as well as the necessary approximations to obtain them. A few examples of such master equations are the Nakajima-Zwanzig equation \cite{nakajima1958, zwanzig1960}, the local-in-time non-Markovian master equation \cite{Breuer_JCmodel, Krovi_2007}, hierarchical equations of motion \cite{HEOM} and more approximate methods such as the Lindblad equation \cite{Lindblad1976, gorini1976, LindbladIntro} (see Ref.~\cite{BRE02} or Ref.~\cite{devega2017} for a comprehensive overview, with the latter focusing on non-Markovian open quantum systems). 
Some of these master equations lead to equivalent results even though their physical interpretation, as well as their derivation may differ.
Additionally, some of these master equations also have equivalent formulations in terms of stochastic unravelings as well as stochastic Schrödinger equations \cite{Carmichael1993Open, gardiner2004handbook}.

In this paper, we focus on the class of Lindblad-type master equations and various unravelings of this type of equation.
The most general Lindblad-type master equation is the canonical local-in-time non-Markovian master equation ($\hbar=1$)
\begin{equation}
    \label{eq:nonmarkovianmaster}
    \Dot{\rho}(t) = -i[H_S(t), \rho(t)] + \mathcal{D}_t(\rho(t)),
\end{equation}
where 
\begin{equation}
    \mathcal{D}_t(\rho) \coloneqq \sum_{l=1}^L\gamma_l(t) \left(A_l(t)\rho A_l^{\dagger}(t)-\frac{1}{2}\left\{A_l^{\dagger}(t) A_l(t), \rho\right\}\right)
\end{equation}
is called the dissipator.
The operator $H_S(t)$ plays the role of the system Hamiltonian \footnote{Note that the environment can induce additional terms in the Hamiltonian and as such $H_S$ generally differ from the Hamiltonian when the system is uncoupled from the environment.}, while the operators $A_l(t)$ are called jump or Lindblad operators with $\gamma_l(t) \in \mathbb{R}$ the jump rates.
Eq.~\eqref{eq:nonmarkovianmaster} is of central interest when studying open systems because of its simple structure and interpretation.

When the timescale of the dynamics of the system is large compared to the timescale of the dynamics of the environment, we may assume that temporal correlations in the bath are $\delta$-like, making the system behave as Markovian quantum system.
In this case, the master equation reduces to the well known and celebrated Lindblad master equation, which differentiates itself from Eq.~\eqref{eq:nonmarkovianmaster} by the additional condition that all the jump rates must satisfy $\gamma_l(t)\geq 0$ at all times, as assumed in the original derivations \cite{Lindblad1976, gorini1976}.
However, for many practical applications, it is not possible to assume that the bath is Markovian. A prominent example where a non-Markovian description is necessary, is the case of superconducting qubits subject to $1/f$  or other pink noises \cite{1_over_f_noise, groszkowski2023}.
Fortunately, for sufficiently small timescales or coupling rates, it is always possible to capture non-Markovian dynamics with a Lindblad-type equation and obtain the local-in-time non-Markovian master equation in Eq.~\eqref{eq:nonmarkovianmaster}, as long as the jump rates $\gamma_l(t)$ are allowed to take negative values \cite{Breuer_THS}.
The significance of the sign in regard to the Markovianity can be understood in terms of information flow: 
when all jump rates have a positive sign, information only flows from the system to the environment.
When the sign of the jump rate is negative, the system can recover information through a backflow of information from the environment into the system  \cite{Negative_Rates_Non-Markovianity}.
We remark that although Eq.~\eqref{eq:nonmarkovianmaster} represents the most general setting, not all of its solutions are physical, i.e., it does not always preserve the defining properties of the density matrix, namely its positivity.
Conditions for the physicality of its solution have been investigated by Hall (2008) \cite{Complete_positivity_Hall_2008}.

For all specific types of  master equations, it is of central interest to find sufficiently fast algorithms to solve them.
Time-local equations, such as the one in  Eq.~\eqref{eq:nonmarkovianmaster},  can be tackled with standard solvers for systems of  linear ordinary differential equations.
This usually involves  the evaluation of the right-hand side of the corresponding equation, which becomes numerically expensive as the dimension $N$ of the Hilbert space increases.
The main contributors to the computational complexity are the matrix-matrix multiplications required to compute the right-hand side of Eq.~\eqref{eq:nonmarkovianmaster}, which by default is of order $\mathcal{O}(N^3)$ \footnote{modern algorithms bring this down to $\mathcal{O}(N^{2.371552})$ \cite{matrix-matrix}}.
When dealing with sufficiently large Hilbert spaces, the simulation of these equations can be sped up using Monte-Carlo methods at the cost of potentially reducing precision \cite{Molmer:93, Piilo_2008}.
The Monte-Carlo Wave-Function (MCWF) approach \cite{Molmer:93} is one such method, relying on the simulation of an ensemble of stochastic state vector evolutions from which the averages of the relevant observables can be constructed. This method is designed to solve the Lindblad-type equation (Eq.~\eqref{eq:nonmarkovianmaster} with $\gamma_l(t)\geq 0$), reducing the computational complexity to $\mathcal{O}(N^2)$. Alas, it does not generalize to the most general case of Eq.~\eqref{eq:nonmarkovianmaster}, where $\gamma_l(t)$ can be negative.
Over the years, multiple methods have been developed to solve Eq.~\eqref{eq:nonmarkovianmaster} using Monte-Carlo methods similar to the MCWF method, such as the doubled Hilbert-space \cite{Breuer_JCmodel}, tripled Hilbert-space \cite{Breuer_THS} and the Non-Markovian Quantum Jump (NMQJ) \cite{Piilo_2008,Piilo_2009} method.
As their name implies, the doubled and tripled Hilbert-space methods extend the size of the Hilbert space, while the NMQJ method modifies the jump probabilities present in the MCWF method to allow for the numerical solution of Eq.~\eqref{eq:nonmarkovianmaster}.
Interestingly, concurrently with providing an iterative method to solve these equations, these methods also provide a formulation of the deterministic equation in form of a stochastic master equation, forming a bridge between both representations.

In this paper, we present the \methodfull{} (\method{}) method, which generalizes the MCWF method in a more natural way compared to the previously mentioned methods. The paper is structured as follows.
First, we give a brief overview of the MCWF method, upon which the \method{} method improves, in Sec.~\ref{sec:mcwf}.
Afterwards, in Sec.~\ref{sec:method} we present and discuss the details of the \method{} method.
In Sec.~\ref{sec:examples} we apply the \method{} method to a few examples to demonstrate its applicability and compare it to the MCWF and NMQJ method.
Finally, we give our concluding remarks in Sec.~\ref{sec:conclusion}. The appendices provide additional details of the derivations.

\section{The Monte-Carlo Wave Function method}
\label{sec:mcwf}

Before we introduce our new method, we first give a brief introduction to the MCWF method.
This sets up the required background knowledge, as well as the context for our method. Moreover, it allows us to focus on the limits of previously developed methods, and provide the theoretical framework to extend the MCWF method.

We start this section with a purely computational motivation to the unraveling of master equations using the Liouville-von Neumann as an example.
We then follow up with a brief overview of the MCWF method, whose aim is to solve Eq.~\eqref{eq:nonmarkovianmaster} in the Markovian case when all rates $\gamma_l(t)$ are non-negative for all times $t$.

\subsection{Unraveling the Liouville-von Neumann equation}
\label{subsec:mcwf_LNE}

As a starting point of the MCWF method, let us consider the Liouville-von Neumann equation
\begin{equation}
    \label{eq:von_neumann}
    \partial_t \rho (t) = -i\comm{H_S(t)}{\rho(t)},
\end{equation}
which can be seen as Eq.~\eqref{eq:nonmarkovianmaster} without dissipator.
Using standard numerical methods, solving this ordinary differential equation involves the evaluation of the right-hand-side of Eq.~\eqref{eq:von_neumann}.
This requires us to perform the matrix-matrix product of $H_S(t)$ and $\rho(t)$ in each iteration step, which is the main contributor to the computational complexity of the method. 
As previously mentioned, given that the dimension of the matrices is $N$, using standard algorithms for matrix-matrix products require on the order of $\mathcal{O}(N^3)$ operations, with more modern algorithms requiring $\mathcal{O}(N^{2.371552})$ operations \cite{matrix-matrix}.

For pure states $\rho(t)=\ket{\psi(t)}\bra{\psi(t)}$, there exists a faster algorithm to solve Eq.~\eqref{eq:von_neumann}, since, in this case,
the Liouville-von Neumann equation is equivalent to the Schrödinger equation
\begin{equation}
    \label{eq:schroedinger}
    \partial_t \ket{\psi(t)} = -iH_S(t)\ket{\psi(t)}.
\end{equation}
The right-hand side of Eq.~\eqref{eq:schroedinger} involves only a matrix-vector product.
This has a computational complexity of $\mathcal{O}(N^2)$, which is asymptotically better than the matrix-matrix product.
Thus, especially for large $N$, it is more efficient to evolve a pure state using the Schrödinger equation rather than the Liouville-von Neumann equation.
This result can even be extended to certain mixed states.
In fact, if the initial density matrix $\rho(t_0)$ at time $t_0$ can be decomposed into a sufficiently small number $M$ of pure states $\ket{\psi_m}$, i.e.,
\begin{equation}
    \label{eq:initial_state_decomposition}
    \rho(t_0) = \sum_{m=1}^M p_m \ketbra{\psi_m},
\end{equation}
with $M \ll N$ and $0 < p_m \le 1$ such that $\sum_{m=1}^M p_m =1$, the computational complexity of solving the Liouville-von Neumann equation by evolving each of the pure states is $\mathcal{O}(M N^2)$.
This computational advantage not only extends to the efficient solution of the density matrix, but also allows us to efficiently obtain expectation values $\langle O\rangle$ of an observable $O$.
An additional advantage to the reduction in computational complexity is that the memory complexity of a solver can be reduced, because it only needs to keep track of the state vector $\ket{\psi(t)}$ rather than the density operator $\rho(t)$ at each time step $t$.
This rewriting of the Liouville-von Neumann equation as a set of Schrödinger equations with different initial conditions is called an unraveling of the equation.

\subsection{Unraveling the Lindblad equation}
\label{subsec:lindbladunravel}
In the previous subsection we have seen that there is an equivalence between the Schrödinger equation and the Liouville-von Neumann equation.

In contrast to the Schrödinger equation, the unraveling of the Lindblad equation is now stochastic in nature, meaning that the evolution of the state vector is probabilistic.
This unraveled equation, which is equivalent to the Lindblad equation, is \cite{BRE02}
\begin{align}
    \label{eq:MCWF_SSE}
    d \ket{\psi(t)} &= \left[-iH_{\text{eff}}(t)-\frac{1}{2}\sum_{l=1}^L \gamma_l\norm{A_l(t)\ket{\psi(t)}}^2\right]\ket{\psi(t)} dt\nonumber\\
    &\hphantom{=}+ \sum_{l=1}^L \left(\frac{A_l(t)\ket{\psi(t)}}{\norm{A_l(t)\ket{\psi(t)}}}-\ket{\psi(t)}\right) dN_l(t),
\end{align}
where $H_{\text{eff}}$ is the effective non-Hermitian Hamiltonian defined as
\begin{equation}
\label{eq:mcwf_effective_hamiltonian}
    H_{\text{eff}}(t) \coloneqq H_S(t)-\frac{i}{2} \sum_{l=1}^L \gamma_l(t) A_l(t)^\dagger A_l(t)
\end{equation}
and $dN_l(t)$ are Poisson increments with expectation value
\begin{equation}
    \langle dN_l(t)\rangle = \gamma_l \norm{A_l(t)\ket{\psi(t)}}^2 dt
\end{equation}
that satisfy
\begin{equation}
    dN_l(t) dN_k(t) = \delta_{lk} dN_l(t).
\end{equation}

Let us start by discussing the individual steps of the MCWF method, as presented by Mølmer et al.\ (1993) \cite{Molmer:93}.
Consider our system to be in the state $\ket{\psi(t)}$ at time $t$.
The state may then be evolved into one of two types of states, a so-called deterministic state
\begin{equation}
\label{eq:mcwf_state_deterministic}
    \ket{\psi_0(t+\delta t)} = \frac{(1-iH_{\text{eff}}(t)\delta t) \ket{\psi(t)}}{\norm{(1-iH_{\text{eff}}(t)\delta t) \ket{\psi(t)}}} \eqqcolon \mathtt{U}_\text{eff}(\ket{\psi(t)})
\end{equation}
or one of the jump states
\begin{equation}
    \label{eq:mcwf_state_jump}
    \ket{\psi_l(t+\delta t)} = \frac{A_l(t)\ket{\psi(t)}}{\norm{A_l(t)\ket{\psi(t)}}} \eqqcolon \mathtt{A}_{l, \mathrm{eff}} (\ket{\psi(t)}),
\end{equation}
with $l=1, \dots, L$. In Eq.~\eqref{eq:mcwf_state_deterministic} and Eq.~\eqref{eq:mcwf_state_jump}, we have introduced, for compactness, the non-linear operators $\mathtt{U}_\text{eff}$ \footnote{Note that, being non-linear, $\mathtt{U}_\text{eff}$ is not a unitary, but we still denote it with a $\mathtt{U}$ since it reduces to a unitary when there is no jump process.} and $\mathtt{A}_{l, \mathrm{eff}}$.  
Each of these evolution targets corresponds to the respective factor of the differentials in Eq.~\eqref{eq:MCWF_SSE}.
The state $\ket{\psi_l(t+\delta t)}$ is chosen with probability
\begin{equation}
    \label{eq:mcwf_probabilities_jump}
   P_l(t) = \delta t \gamma_l(t) \norm{A_l(t)\ket{\psi(t)}},
\end{equation}
while the probability of evolving into the deterministic state is accordingly
\begin{equation}
    \label{eq:mcwf_probabilities_deterministic}
    P_0(t) = 1-\sum_l P_l(t).
\end{equation}
This can be understood as implementing the Poisson increments $d N_l(t)$.

By iterating this random process, we can evolve an initial state $\ket{\psi(t_0)}$ to any time $t>t_0$, and by repeating this process $N$ times we can generate numerous trajectories $\{\ket{\psi_\alpha(t)}\}_{\alpha=1\dots N}$ for the same initial state $\ket{\psi(t_0)}$.
Using a sufficient number of trajectories, it is then possible to compute an approximation to the expectation value of any desired observable $O$ at time $t$ through
\begin{equation}
    \label{eq:expectation_value}
    \langle O (t) \rangle \approx \frac{1}{N}\sum_{\alpha=1}^N \bra{\psi_\alpha(t)} O\ket{\psi_\alpha(t)}.
\end{equation}
Similarly we can compute an approximation $\sigma(t)$ to the density matrix using
\begin{equation}
    \sigma(t) = \frac{1}{N}\sum_{\alpha=1}^N \ketbra{\psi_\alpha(t)}.
\end{equation}
In case we have an initial mixed state $\rho(t_0)$ with decomposition as in Eq.~\eqref{eq:initial_state_decomposition}, we can instead sample the $\ket{\psi_k}$ with probability $p_k$ in the first step, and then perform the stochastic process we have just described. Given that $N\rightarrow \infty$, one can show (\cite{BRE02, Molmer:93, daley2014}) that at any time $t$ this process reproduces, to first order, the density matrix $\rho(t)$, meaning that
\begin{equation}
    \lim_{N\rightarrow\infty} \lim_{\delta t\rightarrow 0}\sigma(t) = \rho(t).
\end{equation}
The number of paths over which to average depends on the desired accuracy for the expectation values and determines whether the Monte-Carlo approach is faster than using the standard numerical approach.
This ultimately depends on the observable we are interested in, and estimates of the number of samples needed to achieve a certain accuracy can be obtained using standard probabilistic inequalities \cite{Molmer:93, daley2014}.

So far, the positivity of each $\gamma_l(t)$ must be ensured.
If one of the jump rates $\gamma_l(t)$ is negative, Eq.~\eqref{eq:mcwf_probabilities_jump} would then lead to a negative jump probability and the MCWF is no longer applicable.

\section{\methodfull{}}
\label{sec:method}
We now present our new NMEP  method, which aims to 
solve Eq.~\eqref{eq:nonmarkovianmaster} in the most general case, thereby generating an unraveling of the equation.
Our method extends the MCWF method to the non-Markovian regime, in which $\gamma_l(t)$ can take negative values.
Following ideas from Refs.~\cite{BRE02, Piilo_2009}, we first reformulate the MCWF method in terms of quantum mechanical ensembles. Given this formalism, we briefly motivate and present our method.
We follow this with a proof and discussion of our method in which we compare it to the NMQJ method of Ref.~ \cite{Piilo_2009}, which also aims to find an unraveling to Eq.~\eqref{eq:nonmarkovianmaster}.

\subsection{Ensemble Interpretation}
\label{subsec:method_ensemble}
To extend the MCWF method discussed in Sec.~\ref{sec:mcwf}, the first step in our method is to reformulate the MCWF method in terms of ensembles.
This rephrasing does not influence the general principle of the MCWF method, but suggests a different practical implementation.

Given a number of samples $N$ and some initial state $\ket{\psi(t_0)}$ of the system, the MCWF method describes the evolution of $N$ copies of the initial state, resulting in at most $N$ different states $\ket{\psi_\alpha(t)}$ at time $t$.
A na\"\i ve implementation iterates over each trajectory, generating them all individually.
This is inefficient, as for each sample starting from $\ket{\psi(t_0)}$, the same matrix-vector products have to be computed many times to reach the states $\ket{\psi_\alpha(t)}$, specifically in the first few steps when most trajectories are still be on the most probable path. 

To efficiently track these repeated matrix-vector products, we record how many times a certain path has been taken and the state of that path.
Essentially, we are keeping track of an ensemble of states and counts that we denote, at each time step, as
\begin{equation}
\label{eq:ensemble}
\{ (\ket{\psi_\alpha(t)},\, N_\alpha(t))\}_{\alpha \in \Gamma(t)},
\end{equation}
where $N_\alpha(t)$ is the state count (or weight) and $\Gamma(t)$ is a set of labels associated with the states in the ensemble, which in general depend on time. We also require the weights $N_\alpha$ to satisfy a normalization condition  
\begin{equation}
\label{eq:norm_cond}
N=\sum_{\alpha \in \Gamma(t)} N_\alpha(t),
\end{equation}
with $N$ the total number of trajectories which remains constant in time.
In this picture, the occurrence of a quantum jump means that the ensemble counts $N_\alpha$ will change, while the evolution of the ensemble members $\ket{\psi_\alpha(t)}$ is obtained according to the rules described in Sec.~\ref{subsec:lindbladunravel}.

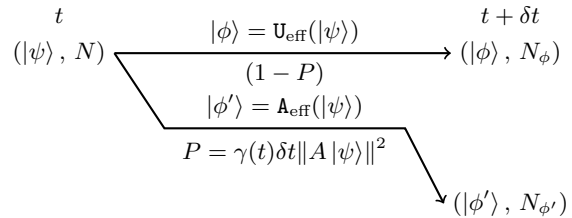
\begin{figure}[h!]
    \centering
    \begin{tikzpicture}
    \path (0,0.5) node(t0) {$t$};
    \path (6,0.5) node(t1) {$t+\delta t$};
    \path (0,0) node(a0) {$(\ket{\psi},\, N)$};
    \path (6,0) node(a2) {$(\ket{\phi},\, N_{\phi})$};
    \path (6,-2) node(a3) {$(\ket{\phi'},\, N_{\phi'})$};
    
    \draw[thick,->](a0.east) -- (1.4,0) -- (4.6,0) node [above, pos=0.5]{$\ket{\phi} = \mathtt{U}_{\textrm{eff}} (\ket{\psi})$} node [below, pos=0.5]{$(1-P)$} -- (a2.west);

    \draw[thick,->](a0.east) -- (1.4,-1) -- (4.6,-1) node [above, pos=0.5]{$\ket{\phi'} = \mathtt{A}_\text{eff} (\ket{\psi})$} node [below, pos=0.5]{$P = \gamma(t)\delta t \norm{A\ket{\psi}}^2$} --(a3.west);

    \end{tikzpicture}
    \caption{A flowchart representing the time evolution of an ensemble with a single ensemble element and a single jump operator using the MCWF method.
    Each arrow represents a possible evolution path for a state in the ensemble. Above each arrow we depict how the state evolves while the probability of evolving accordingly is given below the arrow.
    The ensemble at time $t$ is described by all the ensemble elements below $t$ and accordingly so for the ensemble at time $t+\delta t$.}
    \label{fig:flowchart_MCWF}
\end{figure}

In the case where all trajectories are the same and no quantum jump occurs in any of them up to time $t$, this rewriting is straightforward: at time $t$, our system is in the pure state $\big\{(\ket{\psi(t)},\,N)\big\}$. Considering the case of a single jump operator $A$, during the evolution of the system to the next time step $t+\delta t$, a certain number $N_{\phi'}$ of ensemble elements perform a quantum jump and $N_\phi = N - N_{\phi'}$ perform no quantum jump.
Denoting the deterministic state and the jump state by $\ket{\phi}$ and $\ket{\phi'}$, respectively, the
resulting ensemble is thus $\{(\ket{\phi},\,N_\phi),\, (\ket{\phi'},\, N_{\phi'})\}$.
A corresponding graph of the process is shown in Fig.~\ref{fig:flowchart_MCWF}, where we have used $\mathtt{U}_{\mathrm{eff}}$ to denote the effective evolution described in Eq.~\eqref{eq:mcwf_state_deterministic} and $\mathtt{A}_\text{eff}$ the effective evolution described in Eq.~\eqref{eq:mcwf_state_jump}.

Since we will use the ensemble method of representing a density matrix $\rho(t)$ at time $t$ through an ensemble repeatedly from now on, we explicitly underline the equivalence between the ensemble and the density matrix. Given an ensemble as in Eq.~\eqref{eq:ensemble}, its associated density matrix is
as
\begin{equation}
\label{eq:density_decomposition}
    \rho(t) = \dfrac{1}{N}\sum_{\alpha\in \Gamma(t)} N_\alpha(t)\ketbra{\psi_\alpha(t)}.
\end{equation}
We stress that this decomposition is not unique as there are an infinite number of ways to represent the same density matrix using different ensembles.
As such, there is an equivalence class of ensembles which represent the same density matrix $\rho(t)$.

\subsection{The \method{} method}
\label{subsec:method_method}

Before we introduce the \method{} method, we briefly touch on two approaches which can be used to derive this method.
A more detailed derivation using these approaches is given in Appendix~\ref{section:Appendix_Approaches}.
Both approaches lead to the \method{} method, but they differ in their physical interpretation and require different mathematical insights to derive the \method{} method.

The first approach (see Appendix~\ref{subapp:semiimplicit}) to derive the \method{} method is based on the observation that we can remove the negative sign in the jump probability $P=\gamma\delta t\braket{\phi'}$ when $\gamma<0$ by reversing the sign of the time step $\delta t$.
A step in time with step size $-\delta t$ at time $t +\delta t$ can then be performed using the MCWF formalism.
This ultimately leads to an implicit equation for the evolution of the ensemble which can be solved for explicitly.
In this derivation, many mathematically equivalent variations of the \method{} method arise naturally.
Mathematically, these variations are akin to the implicit formulation of the Euler method.

The second approach (see Appendix~\ref{subapp:direct}) directly reinterprets the jump process occurring in the MCWF method as the creation of a pair of ensemble members with opposite sign.
This pair of ensemble members can be interpreted as the interaction between the system of interest and the environment.

\subsubsection{Algorithm}
\label{subsubsec:algo}

We now go over the details of the \method{} method, whose aim is to solve Eq.~\eqref{eq:nonmarkovianmaster}.
In the language of Eq.~\eqref{eq:ensemble} and Eq.~\ref{eq:density_decomposition}, the \method{} method describes the evolution of 
$\rho(t)\mapsto\rho(t+\delta t)$ as a mapping
\begin{align}
\label{eq:mappingNMEP}
     &\left\{\big(\ket{\psi_\alpha(t)}, N_\alpha(t)\big)\right\}_{\alpha\in \Gamma(t)} \nonumber\\
     &\hspace{5em}\mapsto \left\{\left(\ket{\psi_{\beta}(t+\delta t)},N_{\beta}(t+\delta t)\right)\right\}_{\beta\in \Gamma(t+\delta t)}.
\end{align}

After the ensemble at $t+\delta t$ has been computed, it serves as the starting point for the next iteration of the \method{} method.
In the following, the time argument is set to be arbitrary but fixed and is left out for the ensemble elements.
This allows us to use the following shorthand notation:
$\ket{\psi_\beta}\equiv\ket{\psi_\beta(t+\delta t)}$, $N_\beta \equiv N_\beta(t+\delta t)$, $\ket{\psi_\alpha}\equiv\ket{\psi_\alpha(t)}$, $N_\alpha \equiv N_\alpha(t)$.

The NMEP method computes the new $N_\beta$ and $\ket{\psi_\beta}$ using the following steps:

\begin{enumerate}[label=(\alph*)]
\item Compute the deterministic trajectory $\ket{\psi_{\beta_0}^{(\alpha)}}$ and the jump paths $\ket{\psi_{\beta_l}^{(\alpha)}}$, from each $\ket{\psi_\alpha}$ with $\alpha \in \Gamma(t)$ as follows:
\begin{eqnarray}
\label{eq:method_state_deterministic}
    \ket{\psi_{\beta_0}^{(\alpha)}} &&= \dfrac{(1-i H_{\mathrm{eff}}(t) \delta t)\ket{\psi_\alpha}}{\norm{(1-i H_{\mathrm{eff}}(t) \delta t)\ket{\psi_\alpha}}},\\
\label{eq:method_state_jump}
    \ket{\psi_{\beta_l}^{(\alpha)}} &&= \dfrac{A_l(t) \ket{\psi_\alpha}}{\norm{A_l(t) \ket{\psi_\alpha}}}.
\end{eqnarray}
Here, the non-Hermitian Hamiltonian  $H_{\mathrm{eff}}(t)$ is defined as in Eq.~\eqref{eq:mcwf_effective_hamiltonian}.

\item Let $\mathcal{B}(x,p)$ be a binomial distributed random variable with mean $x$ and success probability $p$. We define the probability $P_{l}^{(\alpha)}$ that a jump with jump operator $A_l(t)$ occurs on the state $\ket{\psi_{\alpha}}$ as
\begin{equation}
    P_{l}^{(\alpha)} = \delta t\, \abs{\gamma_l(t)}\, \norm{A_l(t)\ket{\psi_\alpha}}^2.
\end{equation}
The occupation numbers $N_{\beta_l}^{(\alpha)}$ for the previously computed states $\ket{\psi_{\beta_l}^{(\alpha)}}$ for $l>0$ are randomly sampled as
\begin{eqnarray}\label{eq:jump_ensemble_number}
    N_{\beta_l}^{(\alpha)} && \sim \mathrm{sgn}\,\Big(N_\alpha\gamma_l(t)\Big) \mathcal{B}\left(\abs{N_\alpha},\, P_{l}^{(\alpha)}\right),
\end{eqnarray}
and the occupation of $\ket{\psi_{\beta_0}^{(\alpha)}}$ is computed as
\begin{eqnarray}\label{eq:deterministicoccupation}
    N_{\beta_0}^{(\alpha)} = N_\alpha - \sum_l N_{\beta_l}^{(\alpha)}.
\end{eqnarray}
Eq.~\eqref{eq:deterministicoccupation} connects the occupation number of the deterministic evolution with the jump paths and  ensures that the trace of the density matrix is preserved. We emphasize that the ensemble members can now have a negative ensemble count.
The sign of ensemble members does not influence their propagation, meaning that they propagate using the same rules and may perform quantum jumps in the same way as ensemble members with a positive ensemble count.
The realization of this process is contained within Eq.~\eqref{eq:jump_ensemble_number}.

\item Finally, we can simplify the elements in the ensemble representation of $\rho(t+\delta t)$ if possible.
This means that we remove states whose occupation number is $0$ and combine occupation numbers whose states are equal (up to a phase factor \footnote{In our implementation we remove the phase invariance of states by forcing the first non-zero component to be a positive real number.}). In this way, we can identify new labels $\beta$ associated with the states in the ensemble at time $t + \delta t$ and gather them in a set $\Gamma(t+\delta t)$. This completes the mapping in Eq.~\eqref{eq:mappingNMEP}.
\end{enumerate}

\subsubsection{Proof}
\label{subsubsec:proof}
We now prove that the algorithm described in Sec.~\ref{subsubsec:algo} provides a solution of Eq.~\eqref{eq:nonmarkovianmaster}.
The techniques used in this proof follow closely along the proof of the MCWF \cite{Molmer:93} and NMQJ \cite{Piilo_2009} methods.
We leave details of the steps to Appendix~\ref{section:Appendix_Proof}.

The core idea of the proof is to show that on average a single step of the method satisfies the equation of motion Eq.~\eqref{eq:nonmarkovianmaster}.
Thus, our aim is to compute the average density matrix from the ensemble obtained by performing a single step of the \method{} method given some arbitrary initial state.

The average density matrix $\sigma(t+\delta t)$ is given by all the ensemble elements that are computed from the density matrix at time $t$ in Eq.~\eqref{eq:density_decomposition}.
Thus, we can compute $\sigma(t+\delta t)$ as a mixture of the states that are generated by the \method{} method from the states $\ket{\psi_\alpha}$.
The contributing states to $\sigma(t+\delta t)$ are the states $\ket{\psi_{\beta_0}^{(\alpha)}}$ in Eq.~\eqref{eq:method_state_deterministic}, which represent ``free`` evolution of $\ket{\psi_\alpha}$ and states $\ket{\psi_{\beta_l}^{(\alpha)}}$ with $\ell >0$, which represent an exchange with the bath through the jump operator $A_l$.
Consequently, the average evolution of $\rho(t)$, $\sigma(t+\delta t)$, can be written as
\begin{widetext}
\begin{align}
    \sigma(t+\delta t) = \sum_{\alpha\in \Gamma(t)} \frac{N_\alpha}{N}&\Bigg[\ketbra{\psi_{\beta_0}^{(\alpha)}} + \sum_{l=1}^L P_{l}^{(\alpha)} \mathrm{sgn}\,\Big(\gamma_l(t)\Big) \Bigg(\ketbra{\psi_{\beta_l}^{(\alpha)}}-\ketbra{\psi_{\beta_0}^{(\alpha)}}\Bigg)\Bigg].
\end{align}
\end{widetext}

The sign function $\mathrm{sgn}(.)$ reverses the order of the ensemble member exchange with the bath (see Fig.~\ref{fig:quantum_jumps}) and accommodates the different sign when evolving states that have a negative ensemble count.

After rearranging the projectors and inserting the definition of each state, for the deterministic states we get (see Appendix~\ref{section:Appendix_Proof} for more details)
\begin{widetext}
\begin{align}
    \label{eq:proof_start}
    \ketbra{\psi_{\beta_0}^{(\alpha)}} &= \Big(1-P_{l}^{(\alpha)} \mathrm{sgn}\,\big(\gamma_l(t)\big)\Big)\dfrac{(1-i H_{\mathrm{eff}}(t) \delta t)\ketbra{\psi_\alpha}(1+i H_{\mathrm{eff}}^\dagger(t) \delta t)}{\norm{(1-i H_{\mathrm{eff}}(t) \delta t)\ket{\psi_\alpha}}^2}\\
    \label{eq:proof_end}
    &= \ketbra{\psi_\alpha} -i \delta t [H_S(t),\ketbra{\psi_\alpha}] - \delta t\sum_{l=1}^L \frac{\gamma_l(t)}{2}\Big\{A_l^\dagger(t) A_l(t),\,\ketbra{\psi_\alpha}\Big\} + \mathcal{O}(\delta t^2) && \delta t\rightarrow 0,
\end{align}
and similarly, for the jump state we get 
\begin{align}
    \ketbra{\psi_{\beta_l}^{(\alpha)}} &= P_{l}^{(\alpha)} \mathrm{sgn}\,\Big(\gamma_l(t)\Big) \dfrac{A_l(t) \ketbra{\psi_\alpha}A_l^\dagger(t)}{\norm{A_l(t) \ket{\psi_\alpha}}^2}\\
    &= \gamma_l(t) A_l(t) \ketbra{\psi_\alpha}A_l^\dagger(t).
\end{align}
\end{widetext}

Putting all of this together and inserting the definition for $\rho(t)$ we finally get
\begin{align}
    \sigma(t+\delta t) &= \rho(t) - i\delta t[H_S(t), \rho(t)] + \delta t\mathcal{D}_t(\rho(t))\nonumber\\
    &\phantom{{}=}  + \mathcal{O}(\delta t^2) && \delta t\rightarrow 0,
\end{align}
which is equivalent to
\begin{align}
    \Dot{\rho}(t) = - i\delta t[H_S(t), \rho(t)] + \mathcal{D}_t(\rho(t)) + \mathcal{O}(\delta t) && \delta t\rightarrow 0
\end{align}
if we let $N\rightarrow \infty$.

\subsubsection{Discussion}

As might be evident from the formulation of the new jump process, the \method{} method implicitly defines how to extend Eq.~\eqref{eq:MCWF_SSE} to the non-Markovian case.
In short, the \method{} method inverts the second term in Eq.~\eqref{eq:MCWF_SSE} to
\begin{equation}
    \sum_{l=1}^L \left(\ket{\psi(t)}-\frac{A_l(t)\ket{\psi(t)}}{\norm{A_l(t)\ket{\psi(t)}}}\right) dN_l(t).
\end{equation}
Implementing this as a process requires the addition of new ensemble members, invalidating the assumption of solving this equation using pure trajectories. 

In the case where all $\gamma_l(t)>0$, this method is equivalent to the MCWF method, but expressed in the formalism of quantum mechanical ensembles.
The main difference with the MCWF method is the occurrence of a possible negative sign in Eq.~\eqref{eq:jump_ensemble_number}.
When $\gamma_l(t)<0$, this algorithm effectively coincides with the NMQJ method, whenever the NMQJ method is applicable.
In fact, unlike the NMQJ, our method is even applicable to the most general mathematical context of Eq. \eqref{eq:nonmarkovianmaster}, i.e., the initial condition is some arbitrary Hermitian matrix and Eq.~\eqref{eq:nonmarkovianmaster} is non-physical (i.e., the choice of the operators and their time-dependence is arbitrary).
This improvement over the NMQJ method stems from the fact that the \method{} method does not necessitate appropriate ensemble states to exist within the ensemble to construct the reverse jump operator.
A more in-depth review and comparison with the NMQJ method is provided in Appendix~\ref{section:Appendix_NMQJ}.
We note that in this work we restrict our comparison to the NMQJ method, since this comparison is fairly straightforward.
In fact, the lack of general applicability of the kernel smoothing technique presented in Luoma et al. (2020) \cite{DiffusiveLimitNMQJ} to the NMQJ method is what prompted our initial development of the \method{} method.
Other recent developments (e.g. \cite{RateOperatorUnraveling}) have introduced more complex methods, which aim to solve the same master equation.
While these might be a topic of interest in the future we will refrain from a comparison to them in this work.

If one is interested in determining whether Eq. \eqref{eq:nonmarkovianmaster} is a physical equation, which may not be the case when some approximations were used in its derivation, one may want to add a step to the \method{} method which checks for negative eigenvalues.
We note that this can be formulated as a minimization problem for the Rayleigh quotient $\frac{\bra{\psi}\rho \ket{\psi}}{\braket{\psi}}$ over all $\ket{\psi}$.
The minimization problem can be solved every (few) iteration(s) using an inverse iteration method, making use of the previous smallest eigenvector as the initial guess for the new problem.
A possible matrix-free method for this step is the locally optimal block preconditioned conjugate gradient algorithm \cite{LOBPCG}.

In regard to the computational costs introduced by the \method{} method, we remark that it does add cost in comparison to the MCWF method.
This additional cost presents itself whenever the reverse jump process adds new states to the ensemble, leading to a higher effective number of ensemble states.
In principle the total number of ensemble states may exceed the initial ensemble count $N$.
To make this clear, the case $N=1$ may lead to $2$ or more ensemble members once some jump rate becomes negative.
This additional complexity is only present when $\gamma_l(t)<0$ and plays a less prominent role as $N$ increases. 

As a final remark we note that the \method{} method can be expanded to higher orders in a similar manner as described by Steinbach et al.\ (1995) \cite{higher_order} for the MCWF method.

\section{Examples}
\label{sec:examples}

In this section, we apply the \method{} method to the spin-star model, first discussed in Ref.~\cite{Krovi_2007}, and a model for superconducting transmon qubits presented in Ref.~\cite{SQbit_signatures}.
In Appendix~\ref{subsection:Appendix_NMQJ_c}, we also apply the \method{} method to the Jaynes-Cummings model \cite{JaynesCummingsOriginal, Breuer_JCmodel, BRE02} and compare it to the NMQJ method \cite{Piilo_2009}, showing its applicability in a wider mathematical context.
In the spin-star model we show that the \method{} method is applicable to simple physical equation in which the NMQJ method is not directly applicable. 
In the model for the superconducting transmon we demonstrate the ability to simulate a system in which one of the decay-rates is negative for all times.
Before applying the \method{} method to the models, we briefly introduce them along with their corresponding master equation.

\subsection{Spin-Star Model}
\label{subsec:example_spinstar}

The Ising spin-star system consists of a central coupled to a bath of (non-interacting) spins via $Z Z$-interactions.
Each spin in the bath can have its own independent resonance frequency as well as an independent interaction strength to the central spin.
The reduced system dynamic for the central spin leads to a pure dephasing model, in which only the Pauli $\sigma_z$ operator is present as a jump operator.
Using appropriate system parameters, the jump rate for this operator has regions for which it has a negative sign, making it a good example to apply the \method{} method.

The system Hamiltonian for this system is given by
\begin{equation}
    \label{eq:spin-star-hamiltonian}
    H = \frac{1}{2}\omega_0\sigma^z \otimes \mathds{1}_B + \mathds{1}_S \otimes \sum_{n=1}^N \frac{1}{2}\Omega_n \sigma^z_n + \alpha \sigma^z\otimes\sum_{n=1}^N g_n\sigma_n^z,
\end{equation}
where $\mathds{1}_B$, $\mathds{1}_S$ denote the identity on the bath and the central spin, respectively, $\omega_0/2 \pi$ is the frequency of the central spin, while $\Omega_n/2 \pi$ the frequency of the $n$-th spin in the bath. Additionally, $\alpha >0$ represents a parameter, with dimensions of frequency, that controls the coupling strength of the bath spins to the central spins and $g_n \in [-1, 1]$ are dimensionless parameters associated with each spin in the bath. 
For simplicity, we look at a version of this model, in which the coupling parameters $g_n=g=1$ and the resonance frequencies of the bath $\Omega_n=\Omega$ are equal $\forall n$.
The analytical solution of the reduced system density matrix $\rho_S = \Tr_B{\rho}$ to this problem is derived by Krovi et al.\ (2007) \cite{Krovi_2007}, and, in a frame rotating at frequency $\omega_0$, reads
\begin{equation}
\label{eq:solution_reduced_density_matrix}
    \rho_S(t) = \frac{1}{2} \begin{pmatrix} \rho_{S,11}(t_0) & \rho_{S,12}(t_0)f(t) \\ \rho_{S,21}(t_0)f^*(t) & \rho_{S,22}(t_0)\end{pmatrix},
\end{equation}
where
\begin{equation}
\label{eq:solution_f}
    f(t) = \left(\cos{\left(2 \alpha t\right)}-i\tanh{\left(-\frac{\beta\Omega}{2}\right)}\sin{\left(2 \alpha t\right)}\right)^N.
\end{equation}
This solution assumes a factorizing initial condition $\rho = \rho_S \otimes \rho_B$ in which the state of the bath is a thermal state at inverse temperature $\beta$.
Using the solution as a basis, we can easily compute the associated master equation by taking its derivative, as detailed in Appendix~\ref{section:Appendix_SpinStar}.
The resulting master equation  reads
\begin{equation}
    \partial_t\rho_S(t) = i \left\{\delta(t)\sigma^z,\,\rho_S(t)\right\} + \gamma(t)\left(\sigma^z \rho_S(t) \sigma^z-\rho_S(t) \right)
    \label{eq:spin_star_tcl}
\end{equation}
with the time-dependent Lamb shift
\begin{equation}
\label{eq:lambspinstar}
    \delta(t) = \dfrac{\alpha N \sinh{\left(-\beta\Omega\right)}}{\cos{\left(4 \alpha t\right)+\cosh{\left(-\beta\Omega\right)}}}
\end{equation}
and the time-dependent jump rate
\begin{equation}
\label{eq:decayspinstar}
    \gamma(t) = \dfrac{\alpha N \sin{\left(4 \alpha t\right)}}{\cos{\left(4 \alpha t\right)+\cosh{\left(-\beta\Omega\right)}}}.
\end{equation}

\begin{figure}[t]
    \centering
    \includegraphics[width=0.49\textwidth]{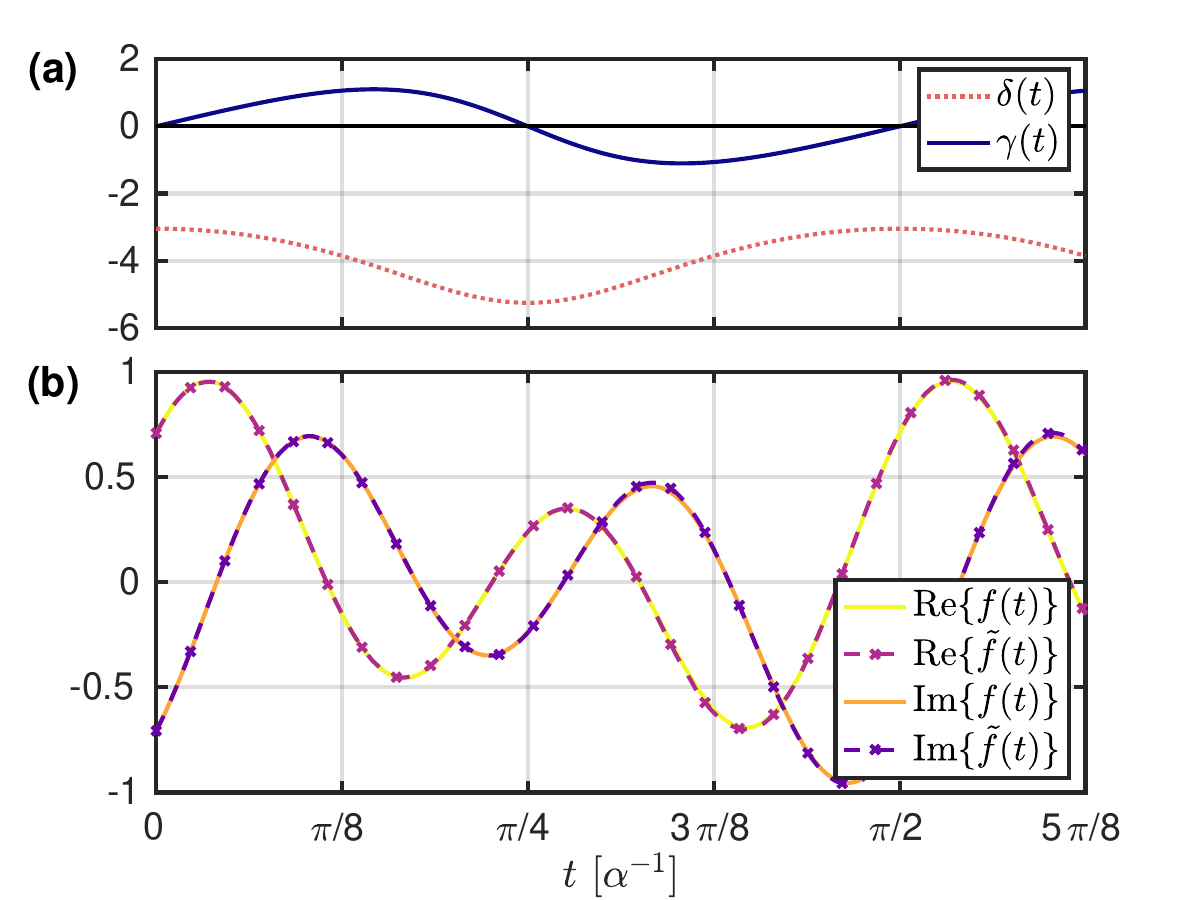}
    \caption{Comparison between the analytical and the numerical results obtained with the \method{} method for the spin-star model. (a) Time dependency of the Lamb-shift $\delta(t)$ and jump rate $\gamma(t)$ of the spin-star model with the system parameters $N=4$ and $\beta\Omega=2$. (b) Analytical result $f(t)$ and simulation result $\Tilde{f}(t)$ of the spin-star model. The simulation parameters are given in the main text.}
    \label{fig:SpinStarComparison}
\end{figure}

We can now apply the \method{} method to Eq.~\eqref{eq:spin_star_tcl} and compare the results directly to the analytic solution of the system given by Eq.~\eqref{eq:solution_reduced_density_matrix}.
We set $\alpha=1$ as the unit of frequency and take the simulation parameters as $t_0=0$, $t_\text{max}=\pi/2+0.5$, $N=4$ \footnote{We chose a small number $N$ of spins in the bath to ensure that the master equation is numerically stable enough for our method to be applicable.}, $\beta\Omega = 2$, $\delta t = 10^{-6} (t_\text{max}-t_0)$, and an initial ensemble given by $\{\frac{1}{\sqrt{2}}\ket{0}+\frac{1+i}{2}\ket{1}, 10^5)\}$.
A plot of the lamb shift $\delta(t)$, the jump rate $\gamma(t)$ as well as a comparison of the resulting analytical and numerical solution can be seen in Fig.~\ref{fig:SpinStarComparison}.

As can be seen in Fig.~\ref{fig:SpinStarComparison}(a), the periodic jump rate assumes negative sign in the region $[\frac{\pi}{4}, \frac{\pi}{2}]$.
This leads to a revival of the coherence, which is visible as an increase of the amplitude of $\abs{\Tilde{f}(t)}$ in that region.
As is clear from Fig.~\ref{fig:SpinStarComparison}, the \method{} method is capable of reproducing the behavior of the solution qualitatively as well as with a sufficient degree of accuracy.
The \method{} method works in presence of both a time-dependent Hamiltonian with Lamb shift $\delta(t)\sigma^z$, and negative and time-dependent jump rates $\gamma(t)$.
It is of interest to mention that, to obtain the results in Fig.~\ref{fig:SpinStarComparison}, we applied a binning step to reduce the number of ensemble members.
In particular, we gathered ensemble members whose state differs by less than $10^{-6}$ in norm. This results in our simulation of the system ending with $50$ ensemble members, severely limiting the number ensemble members that had to be computed.

\subsection{Transmon Qubit Subject to $1/f^{\alpha}$ Noise}
\label{subsec:example_transmon}

\begin{figure}
    \centering
    \includegraphics[width=0.49\textwidth]{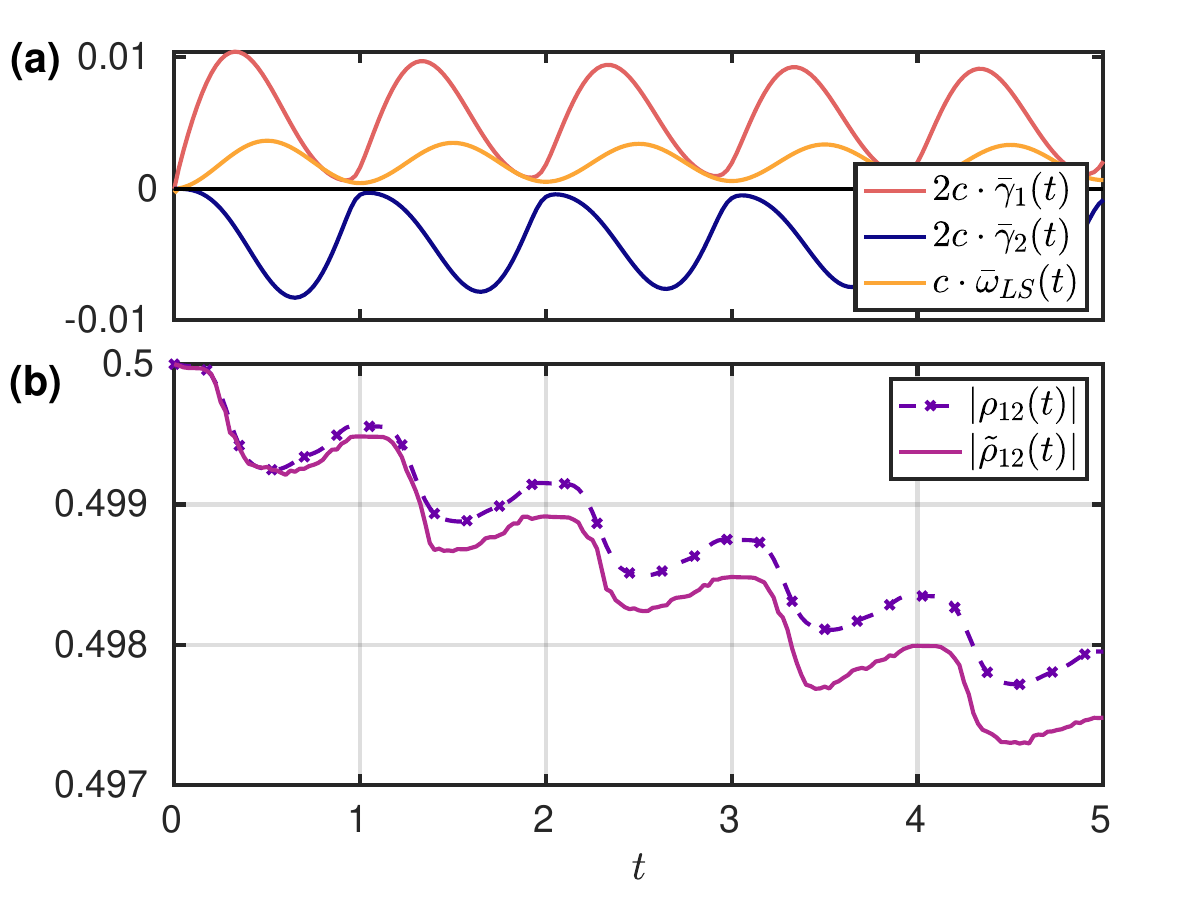}
    \caption{Time evolution for a transmon qubit under $1/f^{\alpha}$ noise. The chosen parameters are detailed in the main text. (a) Lamb-shift and jump rates as a function of time (see Appendix~\ref{section:Appendix_Transmon} for detailed formulas). (b) Absolute value of the coherences in the computational basis as a function of time obtained using the 4th-order Runge-Kutta algorithm ($|\rho_{12}(t)|$) and the \method{} method ($|\tilde{\rho}(_{12}(t))|$).}
    \label{fig:transmon}
\end{figure}

As our second example, we apply the \method{} method to a model of a transmon qubit coupled to a bath of oscillators as it is described by Gulácsi and Burkard (2023) \cite{SQbit_signatures}.
Here, the environment models a Caldeira-Leggett impedance with a $1/f^\alpha$ spectral density, representing the noise introduced by components connected to a transmon qubit. This example is of special interest to us because, as we will see, its associated master equation consists of two jump operators whose associated rates have a fixed opposite sign.
More specifically, while one of the jump operators has a positive sign for all times $t>0$, the other has a negative sign for all times $t>0$ (see Fig.~\ref{fig:transmon}(a)).
An application of previous methods (MCWF, NMQJ) in this setting is not straightforward.

After redefining some terms in Eq. (25) of \cite{SQbit_signatures}, the master equation for this model can be expressed as
\begin{align}\label{eq:me-transmon}
    \dot{\rho}(t) &= i\left[\left(\pi + c \, \Bar{\omega}_\text{LS}(t)\right)\sigma_z,\,\rho(t)\right] + 2 c \sum_{l=1}^2 \Bar{\gamma}_{l}(t) \times \\
    &\hphantom{=} \left(A_l(t)\rho(t) A_l^{\dagger}(t) - \frac{1}{2}\left\{A_l^\dagger(t) A_l(t),\rho(t)\right\}\right).\nonumber
\end{align}
The coefficient $c$ brings this equation into natural units of the system and corresponds to the interaction strength between system and environment.
The Lamb-shift associated with the interaction $\Bar{\omega}_{LS}(t)$ is time-dependent and 
in combination with the time-dependent, negative rates $\Bar{\gamma}_l(t)$, this proves difficult to solve with previous methods.
Both $\Bar{\gamma}_{l}(t)$ and $\Bar{\omega}_{LS}(t)$ are depicted in Fig.~\ref{fig:transmon}(a). We refer to Appendix \ref{section:Appendix_Transmon} for a detailed expression of both terms.
The jump operators $A_l(t)$ in Eq.~\eqref{eq:me-transmon} are the same as in the reference \cite{SQbit_signatures} and are obtained through
\begin{equation}
    \begin{pmatrix}
        A_1(t)\\
        A_2(t)\\
    \end{pmatrix}
    =
    U^T(t)
    \begin{pmatrix}
        \sigma_+\\
        \sigma_-
    \end{pmatrix},
\end{equation}
where $U(t)$ is a unitary matrix that diagonalizes the original coefficient matrix $d_{kl}$ in the reference \cite{SQbit_signatures}.
Again, we refer to Appendix \ref{section:Appendix_Transmon} for the details.
In contrast to the spin-star example discussed in Sec.~\ref{subsec:example_spinstar}, we now have time-dependent jump operators, as well as purely negative rates.
Conceptually, the \method{} method deals with this example the same way as with the previous example.

In the following we set the system parameters as $\alpha=0.9$ and $c=10^{-4}$.
As the simulation parameters for the \method{} method we chose $\delta t = 5\cdot10^{-5}$ and an initial ensemble of $\{\frac{1}{\sqrt{2}}\ket{0}+\frac{1+i}{2}\ket{1}, 10^5)\}$.
Since no exact solution of the system exists, we simulate the evolution of the same system with the classic Runge-Kutta 4th order algorithm with the same simulation parameters.
We use the Runge-Kutta simulation instead of the analytical solution to verify the results of our simulation.
The results of the coherence computed with the \method{} method ($\abs{\Tilde{\rho}_{12}(t)}$) and the Runge-Kutta method ($\abs{\rho_{12}(t)}$) are shown in Fig.~\ref{fig:transmon} (b).
We see that the \method{} method is in qualitative agreement with the Runge-Kutta integration. The quantitative discrepancy is within the expected range, given the different order of the methods and our fairly low number of initial ensemble count of $10^5$, as well as a step size of $5\cdot 10^{-5}$. 
We want to emphasize that solving this particular system using the \method{} method is not useful, since the total number of ensemble members ends up at over 2000 different ensemble members towards the end of the simulation. 
Ideally only 2 ensemble members would be necessary to solve a system with a Hilbert-space dimension of 2.

\section{Conclusion}
\label{sec:conclusion}
With the \method{} method we have presented a natural extension of the MCWF to the non-Markovian regime.
This method builds on the ideas provided by the NMQJ method and introduces a new kind of ensemble, where negative populations are possible.
So far, we have not delved deeply into the consequences of a physical interpretation of this method and as such this remains to be studied in the future.
Similarly to the NMQJ method, the key insight to take away is that for non-Markovian dynamics a pure look at individual trajectories is impossible (see also discussion in Ref.~\cite{wiseman2008}).
Instead, to unravel the non-Markovian master equation correctly, we are forced to look at an ensemble, since new trajectories may branch of from a single existing trajectory.

\section*{Acknowledgements}
We acknowledge support from the Federal Ministry of Education and Research (BMBF) within the framework program “Quantum technologies – from basic research to market”, under the QSolid Project, Grant No. 13N16149.

\appendix 

\section{\method{} - Approaches}
\label{section:Appendix_Approaches}
\subsection{Semi-Implicit approach}
\label{subapp:semiimplicit}
In this section derive the \method{} method by constructing a jump process for the negative rates in Eq. \eqref{eq:nonmarkovianmaster} which is consistent with the MCWF method.
The method we will be using to this end uses the ensemble description as discussed in Sec.~\ref{subsec:method_ensemble}.

We consider the graph given in Fig.~\ref{fig:flowchart_MCWF}, which describes the MCWF method.
Given some initial pure state $\ket{\psi}$ and its ensemble count $N_{\psi}$ at time $t$, the MCWF method computes two states $\ket{\phi}$ and $\ket{\phi'}$ at time $t+\delta t$.
The corresponding ensemble counts of these new states are computed from the previous ensemble counts probabilistically.
The probability to go from $\ket{\psi}$ to $\ket{\phi'}$ is given as $P = \gamma\delta t \braket{\phi'}$.
Our goal is to generate a corresponding graph that describes the evolution of some initial state $\{N_{\psi}, \ket{\psi}\}$ if $\gamma<0$. 

\begin{figure}[h!]
    \centering
    \begin{tikzpicture}
    \path (0,0.5) node(t0) {$t$};
    \path (6,0.5) node(t1) {$t+\delta t$};
    \path (0,0) node(a0) {$(\ket{\psi}, N_{\psi})$};
    \path (0,-2) node(a1) {$(\ket{\psi'}, N_{\psi'})$};
    \path (6,0) node(a2) {$(\ket{\phi}, N_{\phi})$};
    
    \draw[thick,<-](a0.east) -- (1.4,0) -- (4.6,0) node [above, pos=0.5]{$\ket{\psi} = \mathtt{U}_{\textrm{eff}}^{-1} (\ket{\phi})$} node [below, pos=0.5]{$(1-P)$} -- (a2.west);

    \draw[thick,<-](a1.east) -- (1.4,-1) -- (4.6,-1) node [above, pos=0.5]{$\ket{\psi'} = \mathtt{A}_\text{eff}(\ket{\phi})$} node [below, pos=0.5]{$P = \abs{\gamma(\delta t + t)}\delta t \norm{A\ket{\phi}}^2$} --(a2.west);

    \end{tikzpicture}
    \caption{Flowchart representing a step of the MCWF method when applying a reverse time step $-\delta t$ from $t+\delta t$ to $t$ while $\gamma(t+\delta t)<0$.
    This is essentially a mirror image of Fig.~\ref{fig:flowchart_MCWF}. }

    \label{fig:flowchart_NMEP_implicit}
\end{figure}
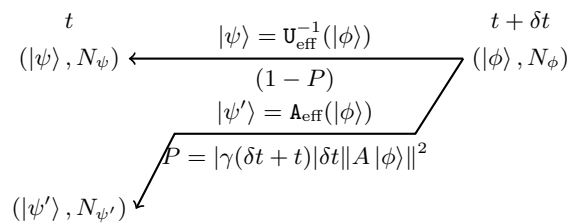

To this end we first make the following observation: if $\gamma<0$ and $\delta t<0$, the probability $P$ does not change its sign.
This suggests that we may perform a reverse step from $t+\delta t$ to $t$ with step size $-\delta t$ to keep the jump probability $P$ positive.
This leads to an implicit formulation of the problem for which we sketch the resulting process in Fig.~\ref{fig:flowchart_NMEP_implicit}.

There are a few mathematical caveats in Fig.~\ref{fig:flowchart_NMEP_implicit} that one needs to pay attention to when comparing it to Fig.~\ref{fig:flowchart_MCWF}.
Overall, these will not change the validity of the method we derive in the following, but requires some strict care when dealing with the mathematical details.
First, the effective evolution operator for the deterministic process $\mathtt{U}_{\textrm{eff}}$ and the jump operator $\mathtt{A}_\text{eff}$ are not exactly the same ones as the ones in Fig.~\ref{fig:flowchart_MCWF}.
This is due to the fact that these operators need to be evaluated at time $t+\delta t$ and not $t$.
We neglect the errors made by this approximation as they are only second order in $\delta t$.
As such it does not matter exactly whether we evaluate $\mathtt{U}_{\textrm{eff}}$ and $\mathtt{A}_\text{eff}$ at $t$ or $t+\delta t$.
Furthermore,  in the reverse step, $\gamma$ needs to be evaluated at $t+\delta t$, however if we may assume that $\gamma$ is sufficiently continuous and can be Taylor expanded to first order this error may be neglected as well.
As such we will leave out the time arguments to keep the notation concise.

With this we have implicitly represented our strictly non-Markovian Monte-Carlo process which moves forward in time as a Markovian process which moves backwards in time.
We are now faced with two problems:
\begin{enumerate}
    \item  We need to determine $N_\phi$ from $N_\psi$.
    \item  Our system starts in the state $\{N_\psi, \ket{\psi}\}$, therefore $N_{\psi'}=0$.
\end{enumerate}

The first problem corresponds to the inversion of a binomial process with probability $1-P$, i.e., determining the number of necessary trials in a binomial process, given that we have had a certain number of successes.
This is a solved problem and the resulting probability distribution is the negative binomial distribution.

\begin{figure}[h!]
    \centering
    \begin{tikzpicture}
    \path (0,0.5) node(t0) {$t$};
    \path (6,0.5) node(t1) {$t+\delta t$};
    \path (0,0) node(a0) {$(\ket{\psi}, N_{\psi})$};
    \path (0,-2) node(a1) {$(\ket{\psi'}, 0)$};
    \path (6,0) node(a2) {$(\ket{\phi}, N_{\phi})$};
    \path (6,-2) node(a3) {$(\ket{\phi'}, N_{\phi'})$};
    
    \draw[thick,<-](a0.east) -- (1.4,0) -- (4.6,0) node [above, pos=0.5]{$\ket{\psi} = \mathtt{U}_{\textrm{eff}}^{-1} (\ket{\phi})$} node [below, pos=0.5]{$(1-P)$} -- (a2.west);

    \draw[thick,<-](a1.east) -- (1.4,-1) -- (4.6,-1) node [above, pos=0.5]{$\ket{\psi'} = \mathtt{A}_\text{eff} (\ket{\phi})$} node [below, pos=0.5]{$P = \abs{\gamma}\delta t \norm{A\ket{\phi}}^2$} --(a2.west);

    \draw[thick,<-](a1.east) -- (1.4,-2) -- (4.6,-2) node [above, pos=0.5]{$\ket{\psi'} = \mathtt{U}_{\textrm{eff}}^{-1} \ket{\phi'}$} node [below, pos=0.5]{} -- (a3.west);
    \end{tikzpicture}
    \caption{Including the boundary conditions into Fig.\ref{fig:flowchart_NMEP_implicit} and taking into account the time evolution of $\ket{\psi'}$, Fig.~\ref{fig:flowchart_NMEP_implicit} leads to an implicit flowchart for the application of the MCWF method to negative rates.} 
    \label{fig:flowchart_NMEP_implicit_new_state}
\end{figure}
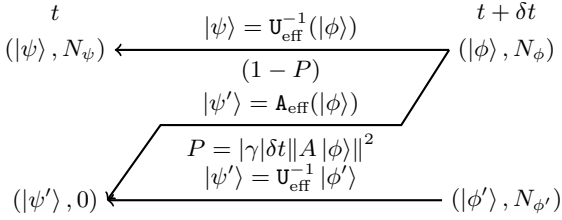

The second problem is a little trickier as we do not know which other states may contribute to $\ket{\psi'}$.
We now make use of the fact that up to first order in $\delta t$ the only other significant contribution to  $\ket{\psi'}$ is through the deterministic evolution of some state $\ket{\phi'}$ that evolves to $\ket{\psi'}$ in the same manner that $\ket{\phi}$ evolves to $\ket{\psi}$.
We sketch the corresponding new graph in Fig.~\ref{fig:flowchart_NMEP_implicit_new_state}.

A priori this may look fairly innocent, however we have hereby introduced the possibility for negative ensemble counts.
To understand how this happens, let us solve the system of equations for the expected evolution of the ensemble counts.
\begin{align}
    \langle N_\psi \rangle = (1-P) \langle N_\phi \rangle\\
    0 = P \langle N_\phi \rangle + \langle N_{\phi'} \rangle 
\end{align}
Keeping only first order terms in $\delta t$ (thus first order in $P$), this may be rearranged to
\begin{align}
    \langle N_\phi \rangle = (1+P) \langle N_\psi \rangle\\
    \langle N_{\phi'} \rangle = -P \langle N_\psi \rangle
\end{align}
or
\begin{align}
    \langle N_{\phi'} \rangle = -P \langle N_\psi \rangle\label{eq:newjumpcount}\\
    \langle N_\phi \rangle = \langle N_\psi \rangle - \langle N_{\phi'} \rangle\label{eq:newdeterministiccount}
\end{align}

\begin{figure}[h!]
    \centering
    \begin{tikzpicture}
    \path (0,0.5) node(t0) {$t$};
    \path (6,0.5) node(t1) {$t+\delta t$};
    \path (0,0) node(a0) {$(\ket{\psi}, N_{\psi})$};
    \path (6,0) node(a2) {$(\ket{\phi}, N_{\phi})$};
    \path (6,-2) node(a3) {$(\ket{\phi'}, N_{\phi'})$};
    
    \draw[thick,->](a0.east) -- (1.4,0) -- (4.6,0) node [above, pos=0.5]{$\ket{\phi} = \mathtt{U}_{\textrm{eff}} (\ket{\psi})$} node [below, pos=0.5]{} -- (a2.west);

    \draw[thick,->](a0.east) -- (1.4,-1) -- (4.6,-1) node [above, pos=0.6]{$\ket{\phi'} = \mathtt{U}_{\textrm{eff}}^{-1} \circ \mathtt{A}_\text{eff} \circ \mathtt{U}_{\textrm{eff}} (\ket{\psi})$} node [below, pos=0.5]{$P = \abs{\gamma}\delta t \norm{A\ket{\phi}}^2$} --(a3.west);

    \end{tikzpicture}
    \caption{After inverting all arrows in Fig.~\ref{fig:flowchart_NMEP_implicit_new_state} and removing the state with $0$ ensemble count from the ensemble at time $t$, this figure now shows an explicit representation of a process that implements the evolution of a system with negative rates.
    An important note is that $P$ no longer represents the probability for $\ket{\psi}$ to evolve towards $\ket{\phi'}$. Instead it represents the probability of $\ket{\psi}$ to induce an ensemble member with negative count to be create in the state $\ket{\phi'}$. Accordingly, $N_\phi=N_\psi -N_{\phi'}$.} 
    \label{fig:flowchart_NMEP_implicit_result}
\end{figure}
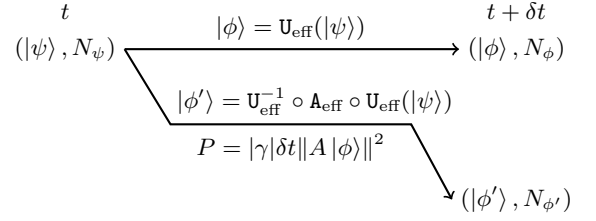

We now draw the final graph for the transition of the state $\ket{\psi}$ to $\ket{\phi}$ and $\ket{\phi'}$, reversing the current process, in Fig.~\ref{fig:flowchart_NMEP_implicit_result}.
The average ensemble counts $\langle N_{\phi'} \rangle$ and $\langle N_{\psi'} \rangle$ have to follow Eq. \eqref{eq:newjumpcount} and Eq. \eqref{eq:newdeterministiccount}, however the ensemble counts given in Fig.~\ref{fig:flowchart_NMEP_implicit_result} are given by a random process.
Here we highlight the fact that $N_{\phi'}$ is negative only when $\gamma(t)<0$.

With this we have essentially derived a method that defines how to create a new ensemble for time $t+\delta t$ given some initial ensemble $\{N_\psi, \ket{\psi}\}$ at time $t$.
We will now simplify a few things to bring the graph in Fig.~\ref{fig:flowchart_NMEP_implicit_result} into a more familiar form.
All the following simplifications are only possible in this way because we want to derive a first order method (in accordance to the MCWF method), ignoring all higher order terms.\\
First, because the process from $\ket{\psi}$ to $\ket{\phi'}$ happens only with probability $\sim \delta t$, we may approximate $\mathtt{U}_{\textrm{eff}}^{-1} \circ \mathtt{A}_\text{eff} \circ \mathtt{U}_{\textrm{eff}} = \mathtt{A}_\text{eff} + \mathcal{O}(\delta t)$ and ignore the linear terms.
Similarly, we may replace $\delta t \norm{A\ket{\phi}}^2$ by $\delta t \norm{A\ket{\psi}}^2$.
Next, instead of computing $N_{\phi'}$ using a negative binomial process, we may just use a binomial process, or any process whose mean is $P$, since we ignore second order contributions.

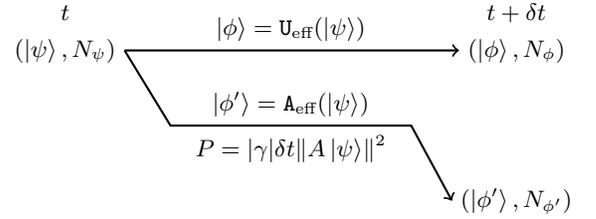
\begin{figure}[h!]
    \centering
    \begin{tikzpicture}
    \path (0,0.5) node(t0) {$t$};
    \path (6,0.5) node(t1) {$t+\delta t$};
    \path (0,0) node(a0) {$(\ket{\psi}, N_{\psi})$};
    \path (6,0) node(a2) {$(\ket{\phi}, N_{\phi})$};
    \path (6,-2) node(a3) {$(\ket{\phi'}, N_{\phi'})$};
    
    \draw[thick,->](a0.east) -- (1.4,0) -- (4.6,0) node [above, pos=0.5]{$\ket{\phi} = \mathtt{U}_{\textrm{eff}} (\ket{\psi})$} node [below, pos=0.5]{} -- (a2.west);

    \draw[thick,->](a0.east) -- (1.4,-1) -- (4.6,-1) node [above, pos=0.5]{$\ket{\phi'} = \mathtt{A}_\text{eff} (\ket{\psi})$} node [below, pos=0.5]{$P = \abs{\gamma}\delta t \norm{A\ket{\psi}}^2$} --(a3.west);

    \end{tikzpicture}
    \caption{Doing some first order approximations, we can simplify Fig.~\ref{fig:flowchart_NMEP_implicit_result} to result in a flowchart that is almost identical to Fig.~\ref{fig:flowchart_MCWF}.} 
    \label{fig:flowchart_NMEP}
\end{figure}

With these considerations we may rewrite the graph in Fig.~\ref{fig:flowchart_NMEP_implicit_result} as in Fig.~\ref{fig:flowchart_NMEP}

With these simplifications the graph in Fig.~\ref{fig:flowchart_NMEP} corresponds exactly to the one in Fig.~\ref{fig:flowchart_MCWF}.
The only difference is the computation of the ensemble counts implied by Eq.~\eqref{eq:newjumpcount}, meaning that a negative jump rate induces a negative population.
Eq.~\eqref{eq:newdeterministiccount} stems from the fact that we still need to enforce conservation of the total number of ensemble members, meaning that
\begin{equation}
    \label{eq:ensemble_number_conservation}
    N_\phi + N_{\phi'} = N_\psi.
\end{equation}
    
To make our algorithm complete, we need to be able to handle initial ensemble members whose ensemble count is negative since $N_{\phi'}$ may be negative.
This is because we still need to evolve the states with a negative ensemble count.
To this end it suffices to invert Eq.~\eqref{eq:newjumpcount} and \eqref{eq:newdeterministiccount} to get
\begin{align}
    \langle -N_{\phi'} \rangle &= -P \langle -N_\psi \rangle\\
    \langle -N_\phi \rangle &= \langle -N_\psi \rangle - \langle -N_{\phi'} \rangle
\end{align}
when $\gamma<0$ or
\begin{align}
    \langle -N_{\phi'} \rangle &= P \langle -N_\psi \rangle\\
    \langle -N_\phi \rangle &= \langle -N_\psi \rangle - \langle -N_{\phi'} \rangle
\end{align}
when $\gamma>0$.

We can now derive the general rule to compute the number of elements $N_{\phi'}$ that perform a quantum jump from $\ket{\psi}$ to $\ket{\phi} = \mathtt{A}_\textrm{eff}(\ket{\psi})$ with arbitrary rate $\gamma(t)$ as
\begin{equation}
    \label{eq:ensemble_count_new}
    N_{\phi'} = \frac{N_\psi\gamma(t)}{\abs{N_\psi\gamma(t)}} \mathcal{B}\left(\abs{N_\psi},\, \delta t\, \abs{\gamma(t)}\, \norm{A\ket{\psi}}^2\right),
\end{equation}
where the fraction may be replaced by the sign function.
The random process $\mathcal{B}(x, p)$ can be any random process mean $x$ and variance $p$, however we will assume $\mathcal{B}$ to be Binomial for practical purposes.
It follows that for ensembles whose ensemble counts $-N_\psi$ is negative, the evolution can be represented as in Fig.~\ref{fig:flowchart_NMEP} by assuming $N_\psi, N_\phi, N_{\phi'}\in\mathbb{Z}$.

We have thus derived, through an implicit Ansatz, how to deal with the negative probability that occurs in the standard MCWF method.
The jump probability in the MCWF has to be adjusted from $P=\gamma \delta t \braket{\phi}$ to $P=\abs{\gamma} \delta t \braket{\phi}$ and the jump process for negative rates $\gamma(t)$ may lead to the creation of a pair of ensemble members with probability $P$: one increases the contribution of $\ket{\phi}$ and one removes from $\ket{\phi'}$.
The states participating in this method remain the same as in the MCWF method.
Computationally, negative jump rates may increase the workload because if $\ket{\phi'}$ does not already exist in the ensemble, which is the generic case, it is added to total number of states that need to be handled.

To generalize from a single jump channel to multiple jump channels requires us to define a jump target and a probability for each jump operator in the same way as for the MCWF and NMQJ method.

\subsection{Direct approach}
\label{subapp:direct}

We now present an alternative way to derive the same method.
The core idea of this Ansatz is to analyze the interaction of the MCWF method and directly postulate an interpretation corresponding to the inverse process associated with the negative rates.
We again consider a system with a single jump operator $A$ and the resulting evolution of some pure state $\ket{\psi}$ at time $t$ to either state $\ket{\phi} = \mathtt{U}_\textrm{eff}(\ket{\psi})$ or $\ket{\phi'} = \mathtt{A}_\text{eff} (\ket{\psi})$ at time $t+\delta t$.
This time, we consider only a single ensemble member to be in the state $\ket{\psi}$.

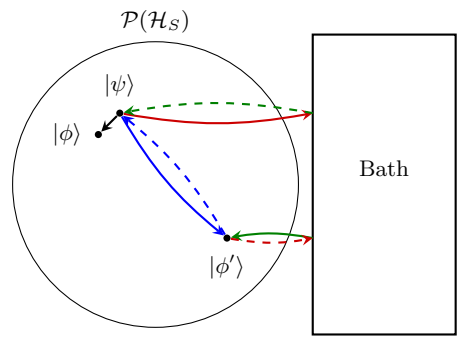
\begin{figure}[ht!]
    \centering
        \centering
        \begin{tikzpicture}[scale=0.95, every node/.style={transform shape}]
            \node[circle, draw, label=above:{$\mathcal{P}(\mathcal{H}_S)$}, minimum size=4cm] at (0.5,0) {};
            \draw[thick] (2.7,2.1) rectangle (4.7,-2.1) node[above, pos=0.5]{Bath};
            
            \node[circle, fill, inner sep=1pt] (A) at (-0.3,0.7) {};
            \node[circle, fill, inner sep=1pt] (B) at (0,1) {};
            \node[circle, fill, inner sep=1pt] (C) at (1.5,-0.75) {};
            \path (4,1) node(Bath1) {};
            \path (4,-1) node(Bath2) {};
    
            \node[label=left:{$\ket{\phi}$}] at (A) {};
            \node[label=above:{$\ket{\psi}$}] at (B) {};
            \node[label=below:{$\ket{\phi'}$}] at (C) {};
            
            \draw[thick, stealth-] (A) -- (B);
            \draw[thick, -stealth, blue] (B) to [bend right=10] (C);
            \draw[thick, dashed, -stealth, blue] (C) to [bend right=10] (B);
            \draw[thick, -stealth, black!20!red] (B) to [bend right=10] (2.7,1) ;
            \draw[thick, dashed, -stealth, black!50!green] (2.7,1) to [bend right=10] (B);
            \draw[thick, -stealth, black!50!green] (2.7,-0.75) to [bend right=10] (C);
            \draw[thick, dashed, -stealth, black!20!red] (C) to [bend right=10] (2.7,-0.75);
        \end{tikzpicture}
        \caption{Interpretation of the jump processes depending on the sign of $\gamma(t)$. The solid lines represent the jump process for $\gamma(t)>0$, which can be seen as the excitation in the state $\ket{\psi}$ leaving the ensemble and the bath returning the corresponding jump state $\ket{\phi'}$. The corresponding jump process for $\gamma(t)<0$ can be seen as reversing this process and is represented using dashed lines. The black line corresponds to the no jump evolution of $\ket{\psi}$ to $\ket{\phi}$.}
    \label{fig:quantum_jumps}
\end{figure}

Each state that participates in the evolution of $\ket{\psi}$ is represented in Fig. \ref{fig:quantum_jumps} as a point in the projective Hilbert-space $\mathcal{P}(\mathcal{H}_S)$.
The distance between $\ket{\psi}$ and $\ket{\phi}$ in the figure has been intentionally chosen to be small to underline the fact that for small $\delta t$ the effective deterministic evolution $\mathtt{U}_\text{eff}$ will be close to the identity, while the jump operator may project the state to any arbitrary state in the Hilbert space.
The transition from $\ket{\psi}$ to $\ket{\phi}$ can be seen as the free evolution of $\ket{\psi}$ given the effective Hamiltonian $H_\text{eff}$.
It is shown in Fig.~\ref{fig:quantum_jumps} as the black transition.
The blue transition in Fig.~\ref{fig:quantum_jumps} represents the jump process connecting $\ket{\psi}$ and the corresponding states $\ket{\phi'}$ after a quantum jump has occurred.

We represent the bath to which our system is coupled with as a box.
We now reinterpret the jump process (blue line) as the excitation in the state $\ket{\psi}$ being absorbed by the bath (red line) and the bath returning this excitation to the ensemble after it has been projected onto the jump state (green line).
In a sense the bath is interacting with the ensemble by exchanging a pair of ensemble member excitation and anti-excitation, with the green line representing an excitation and the red line an anti-excitation that is added to the ensemble of our system.
This corresponds directly to the mathematical formulation of the jump process in Eq.~\eqref{eq:MCWF_SSE}, where the second line denotes that an element $\ket{\psi}$ is removed and replaced by $\mathtt{A}_\textrm{eff}(\ket{\psi})$.
In other words, the quantum jump is effectively the same as the bath taking away the ensemble member we are currently considering and returning the jump state.
This picture is interesting as it allows us to more directly interpret changing jump operators as a change of the bath properties.

We now try to understand what happens when the jump rates become negative.
Intuitively a change in the rate of a process corresponds to a reversal of the process.
This means that the blue line in Fig.~\ref{fig:quantum_jumps} needs to be reversed to represent a process with a negative rate, this is depicted with the dotted blue line.
Using the same logic as before, it is now easy to interpret this reverse process as the bath increasing the ensemble count of $\ket{\psi}$ and decreasing the ensemble count of $\ket{\phi}$, corresponding to the opposite excitation/anti-excitation exchange with the ensemble.
Accordingly, the new change in ensemble members for the negative rate process is pictured using the dotted green and red line.
Going back to Eq.~\eqref{eq:MCWF_SSE}, we can see how the reversal of the process, requires an overall negative sign in front of the Poisson increments.
This leads directly to the process we have just described.

This interpretation is quite convenient as it allows us to eliminate the sign in the jump probability of the MCWF method by moving it to the change in the ensemble members.
Using this interpretation, we get the same method as in the previous derivation. The process happens with the same rate, however, the negative rate jump process has to be reinterpreted as a gain and loss of ensemble members with opposite sign to the positive rate jump process.
This explains the fact that two jump processes, which share the same jump rate and jump operator, but with opposite sign in the jump rate, cancel each other out perfectly, resulting in a purely deterministic trajectory.

While less generally applicable, the advantage of this interpretation is that it is physically more intuitive than the previous one, which required a time reversed process and the solution of an implicit system.
Unfortunately, because we are dealing with a time-local equation, it is highly non-physical for pure states.
In principle, the average evolution will not lead to non-physical ensembles in their entirety, however, single trajectories can no longer be associated to a classical probability of occurrence in an ensemble, since they may split into multiple trajectories.

\section{Proof - Details}
\label{section:Appendix_Proof}
In this appendix, we show explicitly the missing step in the proof given in Sec.~\ref{subsubsec:proof}, that is the step from Eq. \eqref{eq:proof_start} to \eqref{eq:proof_end}.

We start by computing the numerator in Eq.~\eqref{eq:proof_start}
\begin{widetext}
\begin{align}
    (1-i\delta t H_{\mathrm{eff}}(t))\ketbra{\psi_\alpha}(1-i\delta t H_{\mathrm{eff}}(t))^\dagger &= \ketbra{\psi_\alpha} - i\delta t \Big[H_S(t), \ketbra{\psi_\alpha}\Big] \nonumber\\
    &  - \frac{\delta t}{2}\sum_{l=1}^L \gamma_l(t) \Big\{A_l^\dagger(t)A_l(t),\ketbra{\psi_\alpha}\Big\} + \mathcal{O}(\delta t^2) && \delta t\rightarrow 0.
\end{align}
After this we compute the remainder of the fraction through Taylor expansion as
\begin{align}
    \frac{1}{\norm{(1-i\delta t H_\text{eff}(t))\ket{\psi_\alpha}}^2} &= 1 + \delta t \sum_{l=1}^L \gamma_l(t) \bra{\psi_\alpha}A_l^\dagger(t) A_l(t)\ket{\psi_\alpha} + \mathcal{O}(\delta t^2) && \delta t\rightarrow 0,
\end{align}
and multiplying with the remaining term results in
\begin{align}
    \frac{\Big(1-\sum_{l=1}^L P_{l}^{(\alpha)} \mathrm{sgn}\,\big(\gamma_l(t)\big)\Big)}{\norm{(1-i\delta t H_\text{eff}(t))\ket{\psi_\alpha}}^2} &= 1+\mathcal{O}(\delta t^2) && \delta t\rightarrow 0.
\end{align}
\end{widetext}
Multiplying everything together and rearranging gives the desired result from Eq. \eqref{eq:proof_end}.

\section{Non-Markovian Quantum Jumps}
\label{section:Appendix_NMQJ}
In this appendix we review the non-Markovian Quantum Jumps (NMQJ) method \cite{Piilo_2008}, whose aim is the same as the \method{} method, and highlight the limitations of the method when compared to the \method{} method.

\subsection{NMQJ method}

We have seen that, in terms of ensemble members, the MCWF method extends the current ensemble with new states and redistributes the ensemble counts in a stochastic manner.
The central ideas of the NMQJ method are 1.\ to realize that a negative rate of some process should be equivalent to the inverse process with a positive rate and 2.\ to find an appropriate inverse process and transition probability to the one described by the MCWF method.

Naively, an inverse process to the MCWF method corresponds to finding some reverse jumps which implement the transition (using our previous notation) from $\{(\ket{\phi},\,N_\phi),\, (\ket{\phi'},\, N_{\phi'})\}$ to $\big\{(\ket{\psi},\,N_\psi)\big\}$.
As we can already see, this process requires the knowledge of multiple different ensemble members in the current time step and the relationship between these is related to the jump channel.

The NMQJ method now proposes the following algorithm for the quantum jumps over negative channels:
\begin{itemize}
    \item [1.] Given some ensemble $\big\{(\ket{\psi_\alpha},\,N_\alpha)\big\}_{\alpha\in \Gamma(t)}$ at time $t$, for each negative channel $A_l(t)$, a reverse jump can be performed from $\ket{\psi_\alpha}$ to $\ket{\psi_{\alpha'}}$, with $\alpha, \alpha'\in\ \Gamma(t)$, if 
    
    \begin{equation}
        \label{eq:nmqj_reverse_jump_condition}
        \ket{\psi_\alpha} = \frac{A_l(t)\ket{\psi_{\alpha'}}}{\norm{A_l(t)\ket{\psi_{\alpha'}}}}
    \end{equation}.
    \item [2.] The probability for this quantum jump to happen in the step from $t$ to $t+\delta t$ is given by
    \begin{equation}
        \label{eq:nmqj_reverse_jump_probability}
        P_{\alpha\rightarrow\alpha'} = \frac{N_{\alpha'}}{N_\alpha} \abs{\gamma_l(t)} \delta t \norm{\ket{\psi_\alpha}}
    \end{equation}
\end{itemize}

\begin{figure}[h!]
    \centering
    \begin{tikzpicture}
    \path (0,0.5) node(t0) {$t$};
    \path (6,0.5) node(t1) {$t+\delta t$};
    \path (0,0) node(a0) {$(\ket{\psi_\alpha},\, N_{\alpha})$};
    \path (0,-2) node(a1) {$(\ket{\psi_{\alpha'}},\, N_{\alpha'})$};
    \path (6,0) node(a2) {$(\ket{\phi},\, N_{\phi})$};
    \path (6,-2) node(a3) {$(\ket{\psi_{\alpha'}},\, N_{\phi'})$};
    
    \draw[thick,->](a0.east) -- (1.4,0) -- (4.6,0) node [above, pos=0.5]{$\ket{\phi} = \mathtt{U}_{\textrm{eff}} (\ket{\psi_{\alpha}})$} node [below, pos=0.5]{$(1-P_{\alpha\rightarrow\alpha'})$} -- (a2.west);

    \draw[thick,->](a0.east) -- (1.4,-1) -- (4.6,-1) node [above, pos=0.5]{$\ket{\psi_{\alpha'}} = \ketbra{\psi_{\alpha'}}{\psi_{\alpha}} \ket{\psi_{\alpha}}$} node [below, pos=0.5]{$P_{\alpha\rightarrow\alpha'}$} --(a3.west);
    \end{tikzpicture}
    \caption{Flowchart representing the NMQJ method. Given the ensemble state at time $t$ on the left, the state $\ket{\psi_\alpha}$ can perform reverse jumps to $\ket{\psi_{\alpha'}}$ with probability $P_{\alpha\rightarrow\alpha'}$ as given in Eq.~\eqref{eq:nmqj_reverse_jump_probability}. The non-jump evolution is the same as in the MCWF method.
    In contrast to Fig.\ref{fig:flowchart_NMEP}, the probability of a reverse jump to happen depends on the makeup of the ensemble at time $t$.}
    \label{fig:flowchart_NMQJ}
\end{figure}
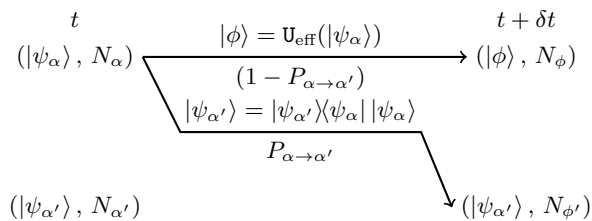

A graph corresponding to the reverse quantum jump process is shown in Fig~\ref{fig:flowchart_NMQJ}.
Note that even though the state $\ket{\psi_{\alpha'}}$ is present in the second ensemble at time $t+\delta t$, it is not directly linked to the one in the first ensemble at time $t$.
This means that in principle these states of the ensembles are to be taken independent of each other.
As we will see, we may replace $\ket{\psi_{\alpha'}}$ in the ensemble at $t+\delta t$ by $\mathtt{U}_\mathrm{eff}\ket{\psi_\alpha'}$ without introducing much more error than already present in the method.
This means $\ket{\psi_\alpha}$ effectively jumps to whichever state $\ket{\psi_\alpha'}$ evolves to deterministically.

\subsection{Limitations}

Let us now discuss some limitations and problems to the NMQJ method.
We start this section with the most glaring issue of the NMQJ method, which is the fact that the only way a jump can happen is if the corresponding state exists within the ensemble. In fact, if it does not exist, it can be interpreted as having $0$ ensemble count, making the  probability for such a jump undefined because it tends to infinity.
The authors of the NMQJ method determine that when their algorithm approaches this region it has to terminate.

From a purely algorithmic perspective, this is the only problem, however there are a few more issues with this method which stem from an implementation point of view.

First, the previous problem is reinforced by the fact that near misses where $\ket{\psi_\alpha} \approx \frac{A_j(t)\ket{\psi_{\alpha'}}}{\norm{A_j(t)\ket{\psi_{\alpha'}}}}$ but not quite equal, which can already happen because of numerical inaccuracies, can make the application of the method difficult.
This can be resolved by an additional binning step, making the Hilbert-space coarse grained, which, however, comes with additional complexity.

Furthermore, the method is now no longer easily parallelizable since the method requires each member of the ensemble to know about the others.
This is in fact the main downside compared to the MCWF method, where each trajectory can be computed independently of the others.

We highlight the fact that because of these issues the NMQJ method is especially unsuited to solve the general non-Markovian master equation (Eq.~ \eqref{eq:nonmarkovianmaster}) unless it presents very specific symmetries which allow us for an easy treatment of the ensemble

\subsection{Comparison to \method{} method}
\label{subsection:Appendix_NMQJ_c}

\begin{figure}
    \centering
    \includegraphics[width=0.49\textwidth]{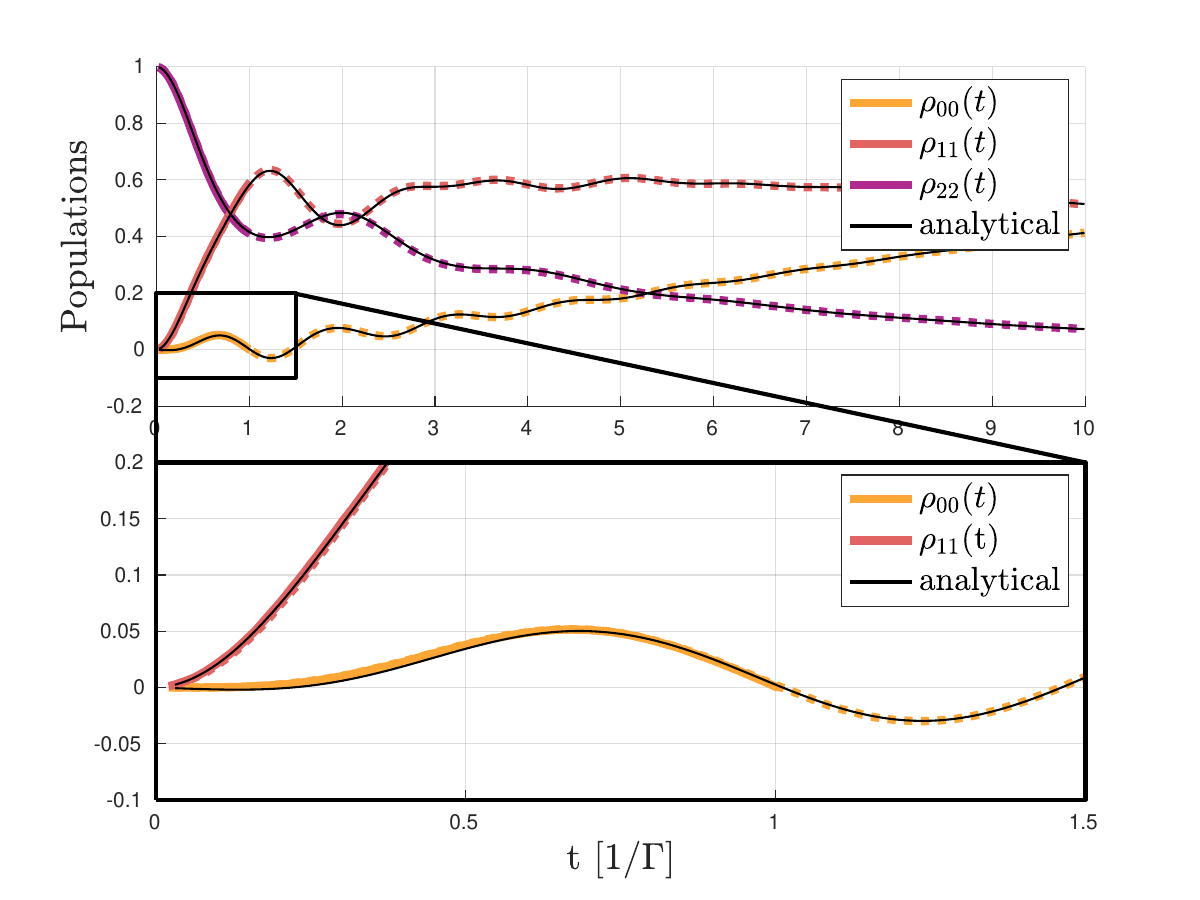}
    \caption{Comparison between the NMQJ and \method{} method using the same example as the one used in Figure 9 of Piilo et al.\ (2009) \cite{Piilo_2009}. At $t\approx 1$, the NMQJ method (solid line) stops being applicable once the master equation produces an unphysical result\footnote{This can happen even when the result is still physical.}. The \method{} method (dashed line) produces a result for all times, solving the master equation in the mathematically most general sense (analytical solution given in black).}
    \label{fig:NMEP_JaynesCummings}
\end{figure}

We now compare the NMQJ method to the \method{} method and demonstrate the usefulness of the latter in the general mathematical context of Eq.~\eqref{eq:nonmarkovianmaster}.
In Piilo et al.\ (2009) \cite{Piilo_2009} the NMQJ is applied to a generalization of the Jaynes-Cummings model \cite{JaynesCummingsOriginal, Breuer_JCmodel, Piilo_2009, BRE02}.
To demonstrate the generality the \method{} method we apply it to the three level ladder system presented by Piilo et al.\ (2009) \cite{Piilo_2009} and compare our results to the ones obtained using the NMQJ method.
To apply the \method{} method, we use the same settings as in section IV-C-4 of Piilo et al.\ (2009) \cite{Piilo_2009} and reproduce their results shown in Figure 9 of that paper, the results are shown in Fig.~\ref{fig:NMEP_JaynesCummings}.

In this example, the master equation derived by generalizing the Jaynes-Cummings model is an approximation of the exact master equation.
The consequence of this is that the solution of this master equation does not produce a valid density operator for all initial conditions and evolution times, meaning that the master equation does not produce physical solutions for the density operator.
More specifically, one of the eigenvalues of the density operator becomes negative given the right initial condition.
In the application of the NMQJ method to this example, this causes the method to stop as soon as the ensemble is unable to represent the state of the system, when it would need ensemble members with a negative ensemble count to account for the negative eigenvalue of the density operator.
Fortunately, because of the structure of the problem, it is easy to identify this negativity of the density matrix with the loss of physicality of the master equation, but this is not true in general.
In the case of the \method{} method, the point at which the system loses physicality does not impact the application of the method, as it produces a solution for all times.
The ensemble generated by the NMQJ and the \method{} method behave in the same way, except for the difference produced by the factor $N_{\alpha'}/N_\alpha$ in the jump rate.
The point at which the solution loses physicality can still be determined in the same way as in the NMQJ method. However, a different method using an estimator for the lowest eigenvalue may be more practical in general.

\section{Spin-Star - Derivation}
\label{section:Appendix_SpinStar}
To derive the master equation governing the Ising spin-star system we take the derivative of $\rho_S(t)$ defined in Eq.~\eqref{eq:solution_reduced_density_matrix} w.r.t. $t$ and try to relate the result to $\rho_S(t)$.
\begin{align}
    \partial_t \rho_S(t) &= \partial_t \frac{1}{2} \begin{pmatrix} {\rho_S}_{11}(0) & {\rho_S}_{12}(0)f(t) \\ {\rho_S}_{21}(0)f^*(t) & {\rho_S}_{22}(0)\end{pmatrix} \nonumber \\
    &= \frac{1}{2} \begin{pmatrix} 0 & {\rho_S}_{12}(0)\partial_t f(t) \\ {\rho_S}_{21}(0)\partial_t f^*(t) & 0 \end{pmatrix} \label{eq:spin-star-derivation-1}.
\end{align}
Using the definition of $f(t)$ in Eq.~\eqref{eq:solution_f}
\begin{align}
    \partial_t f(t) &= \underbrace{\dfrac{2\alpha N\left(-\sin{\left(2 \alpha t\right)}-i\tanh{\left(-\frac{\beta\Omega}{2}\right)}\cos{\left(2 \alpha t\right)}\right)}{\left(\cos{\left(2 \alpha t\right)}-i\tanh{\left(-\frac{\beta\Omega}{2}\right)}\sin{\left(2 \alpha t\right)}\right)}}_{\eqqcolon -4 \zeta(t)} f(t) \nonumber \\
    &= 2\left(-2\Re{\zeta(t)} -2i\Im{\zeta(t)} \right) f(t).
\end{align}
One may verify that defining $\delta(t) \coloneqq \Im{\zeta(t)}$ and $\gamma(t) \coloneqq \Re{\zeta(t)}$ gives Eq.~\eqref{eq:lambspinstar} and Eq.~\eqref{eq:decayspinstar} in the main text, respectively. Replacing the result for $\partial_t f(t)$ in Eq.~\eqref{eq:spin-star-derivation-1}, we find Eq.~\eqref{eq:spin_star_tcl} in the main text, which is the desired master equation.

\section{Transmon - Details}
\label{section:Appendix_Transmon}
Here we give a more detailed expression for the coefficients $\Tilde{\gamma}_k(t)$ and $\Tilde{\omega}_{LS}(t)$ as well as the unitary matrix $U(t)$.
The starting point is Eq.~(21) in Piilo et al.\ (2009) \cite{SQbit_signatures}
\begin{align}
    \label{eq:transmon_me_source}
    \Dot{\rho}(t) &= -i [H(t), \rho(t)] \nonumber\\
    &\hphantom{=}+ \eta^2 \sum_{k,l=\pm} d_{kl}(t)\left(\sigma_k\rho(t)\sigma_l^\dagger - \frac{1}{2}\{\sigma_l^\dagger\sigma_k,\rho(t)\}\right).
\end{align}
which is brought into the following form
\begin{align}\label{eq:born-me-dimless}
    \frac{\partial \rho}{\partial s}(s) &= i\left[\left(\pi + c\Bar{\omega}_\text{LS}(s)\right)\sigma_z,\,\rho(s)\right] \nonumber\\
    &\hphantom{=}+ \sum_{k,l=\pm} 2c \Bar{d}_{kl}(s)\left(\sigma_k\rho(s)\sigma_l^\dagger - \frac{1}{2}\left\{\sigma_l^\dagger\sigma_k,\rho(s)\right\}\right)
\end{align}
using the following substitutions
\begin{align}
    c \coloneqq \frac{Ae^2\eta^2}{\omega_q^{\alpha+1}},\\
    s = \frac{\omega_q t}{2\pi}.
\end{align}

The modified coefficient matrix $(\Bar{d}_{kl}(t))_{k;l}$ then becomes
\begin{align}
    &(\Bar{d}_{kl}(s))_{k;l} = \nonumber\\
    &\begin{pmatrix}
        \Bar{\gamma}_+(s) & -\frac{\Bar{\gamma}_+(s) + \Bar{\gamma}_-(s)}{2} - i\Bar{\omega}_\text{LS}(s)\\
        -\frac{\Bar{\gamma}_+(s) + \Bar{\gamma}_-(s)}{2} + i\Bar{\omega}_\text{LS}(s) & \Bar{\gamma}_-(s)
    \end{pmatrix}
\end{align}
where
\begin{align}
    &\Bar{\gamma}_\pm(s)\coloneqq\nonumber\\
    & 2\frac{\Gamma{(\alpha-1)}}{\alpha}\left(\sin{\left(\frac{\pi\alpha}{2}\right)}f_{\cos}(2\pi s)\pm
    \cos{\left(\frac{\pi\alpha}{2}\right)} f_{\sin}(2\pi s)\right)
\end{align}
and
\begin{align}
    \Bar{\omega}_\text{LS}(s) &\coloneqq 2\frac{\Gamma{(\alpha-1)}}{\alpha} \sin{\left(\frac{\pi\alpha}{2}\right)} f_{\sin}(2\pi s).
\end{align}

The functions $f_\text{cos}$ and $f_\text{sin}$ depend on the non-analytically solvable integrals
\begin{align}
    f_{\cos}(2\pi s) &\coloneqq \int_0^{(2\pi s)^\alpha}\cos{\left(u^{\frac{1}{\alpha}}\right)}~\dd u\\
    \label{eq:f_sin_u}
    f_{\sin}(2\pi s) &\coloneqq\int_0^{(2\pi s)^\alpha}\sin{\left(u^{\frac{1}{\alpha}}\right)}~\dd u
\end{align}
and are computed numerically once at the beginning of the simulation.
We can now obtain $U(t)$ through diagonalization of the coefficient matrix $(\Bar{d}_{kl}(s))_{k;l}$
\begin{equation}
    (\Bar{d}_{kl}(s))_{k;l} = U(t) D(s) U^\dagger(t).
\end{equation}

This is in each time step during the simulation.
The diagonal matrix $D(s)$ contains the jump rates $\Tilde{\gamma}_{1/2}$
\begin{equation}
    D(s) = \begin{pmatrix}
        \Tilde{\gamma}_1 & 0\\
        0 & \Tilde{\gamma}_2
    \end{pmatrix}.
\end{equation}

We note that the jump rates can be computed directly by solving the characteristic polynomial.

\bibliography{bibliography}
\end{document}